\documentclass[twocolumn,twocolumn]{IEEEtran}

\usepackage[T1]{fontenc}
\usepackage{color}
\usepackage{array}
\usepackage{float}
\usepackage{mathrsfs}
\usepackage{amsmath}
\usepackage{amssymb}
\usepackage{graphicx}

\usepackage{fancyhdr}

\usepackage[unicode=true,
 bookmarks=true,bookmarksnumbered=true,bookmarksopen=true,bookmarksopenlevel=1,
 breaklinks=false,pdfborder={0 0 0},pdfborderstyle={},backref=false,colorlinks=false]
 {hyperref}
\hypersetup{pdftitle={Your Title},
 pdfauthor={Your Name},
 pdfpagelayout=OneColumn, pdfnewwindow=true, pdfstartview=XYZ, plainpages=false}

\makeatletter

\floatstyle{ruled}
\newfloat{algorithm}{tbp}{loa}
\providecommand{\algorithmname}{Algorithm}
\floatname{algorithm}{\protect\algorithmname}

\usepackage[caption=false,font=footnotesize]{subfig}
\usepackage{algorithm}
\usepackage{algorithmic}

\usepackage{multirow} 
\usepackage{amsmath} 
\usepackage{xcolor}

\allowdisplaybreaks[4]

\ifCLASSOPTIONcompsoc
\usepackage[caption=false,font=normalsize,labelfont=sf,textfont=sf]{subfig}
\else
\usepackage[caption=false,font=footnotesize]{subfig}
\fi

\usepackage{booktabs}

\usepackage{cite}
\usepackage{bm}
\usepackage{algorithmic}
\usepackage{algorithm}
\usepackage{graphicx}
\IEEEoverridecommandlockouts

\usepackage{lettrine}

\usepackage{geometry}
\@ifundefined{showcaptionsetup}{}{%
 \PassOptionsToPackage{caption=false}{subfig}}
\usepackage{subfig}
\makeatother

\usepackage{cases}

\begin{document}
\hypersetup{hidelinks}

\title{\textcolor{black}{Service Function Chain Dynamic Scheduling in Space-Air-Ground Integrated Networks}}

\author{\IEEEauthorblockN{Ziye Jia, \IEEEmembership{Member, IEEE,} Yilu Cao, Lijun He,  \IEEEmembership{Member, IEEE, }Qihui Wu, \IEEEmembership{Fellow, IEEE, }\\Qiuming Zhu, \IEEEmembership{Senior Member, IEEE, }Dusit Niyato, \IEEEmembership{Fellow, IEEE, }and Zhu Han, \IEEEmembership{Fellow, IEEE}}
\thanks{Copyright (c) 2025 IEEE. Personal use of this material is permitted. However, permission to use this material for any other purposes must be obtained from the IEEE by sending a request to pubs-permissions@ieee.org.}
\thanks{Ziye Jia is with the College of Electronic and Information Engineering, Nanjing University of Aeronautics and Astronautics, Nanjing 211106, China, and also with the State Key Laboratory of ISN, Xidian University, Xian 710071, China (e-mail: jiaziye@nuaa.edu.cn).}
\thanks{Yilu Cao, Qihui Wu and Qiuming Zhu are with the College of Electronic and Information Engineering, Nanjing University of Aeronautics and Astronautics, Nanjing 211106, China (e-mail: caoyilu@nuaa.edu.cn; wuqihui@nuaa.edu.cn; zhuqiuming@nuaa.edu.cn).}
\thanks{Lijun He is with the School of Information and Control Engineering, China University of Mining and Technology, Xuzhou 221116, China (e-mail: lijunhe@cumt.edu.cn).}
\thanks{Dusit Niyato is with the School of Computer Science and Engineering, Nanyang Technological University, Singapore 639798 (e-mail: dniyato@ntu.edu.sg).}
\thanks{Zhu Han is with the Department of Electrical and Computer Engineering at the University of Houston, Houston, TX 77004 USA, and also with the Department of Computer Science and Engineering, Kyung Hee University, Seoul, South Korea, 446-701 (e-mail: hanzhu22@gmail.com).}}

\maketitle

\fancyhf{}

\fancyhead[R]{\ifodd\value{page}\scriptsize\thepage\else\fi}
\fancyhead[L]{\ifodd\value{page}\else\scriptsize\thepage\fi}

\renewcommand{\headrulewidth}{0pt}

\pagestyle{fancy}

\thispagestyle{fancy}
\begin{abstract}
    As an important component of the sixth generation communication technologies, the space-air-ground integrated network (SAGIN) attracts increasing attentions in recent years. However, due to the mobility and heterogeneity of the components such as satellites and unmanned aerial vehicles in multi-layer SAGIN, the challenges of inefficient resource allocation and management complexity are aggregated. To this end, the network function virtualization technology is introduced and can be implemented via service function chains (SFCs) deployment. However, urgent unexpected tasks may bring conflicts and resource competition during SFC deployment, and how to schedule the SFCs of multiple tasks in SAGIN is a key issue. In this paper, we address the dynamic and complexity of SAGIN by presenting a reconfigurable time extension graph and further propose the dynamic SFC scheduling model. Then, we formulate the SFC scheduling problem to maximize the number of successful deployed SFCs within limited resources and time horizons. Since the problem is in the form of integer linear programming and intractable to solve, we propose the algorithm by incorporating deep reinforcement learning. Finally, simulation results show that the proposed algorithm has better convergence and performance compared to other benchmark algorithms.
    
\end{abstract}

\newcommand{\CLASSINPUTtoptextmargin}{0.8in}

\newcommand{\CLASSINPUTbottomtextmargin}{1in}

\section{Introduction}

\lettrine[lines=2]{T}{he} space-air-ground integrated networks (SAGINs) stand out as pivotal elements in the evolution of the sixth generation communication technologies in recent years. SAGIN can provide global services, especially for the remote areas, which gathers significant attentions from both academia and industry \cite{8368236}, \cite{cheng20226g}. SAGIN is mainly composed of satellites, unmanned aerial vehicles (UAVs), ground stations, as well as various users \cite{9722775}. Compared with terrestrial networks, SAGIN is a multi-layer heterogeneous network with diverse resources and complex structure, which can support various tasks for global coverage \cite{10440193}. However, in SAGIN, satellites move periodically with high dynamic, while UAVs move flexibly with detailed planning. Besides, the resource capabilities of diverse nodes are different \cite{10244089}. Since the traditional satellites or aerial UAVs are generally designed for certain types of tasks \cite{8612450}, the different layers of networks are isolated and cannot share resources, resulting in low resource utilization, large overhead, and unsatisfied services. Therefore, it is necessary to cooperate the multiple resources in SAGIN to provide better services for the increasing number of terrestrial users.

The network function virtualization (NFV) technology can be introduced in SAGIN to improve the resource management efficiency. In particular, NFV decouples the network functions from hardware devices by deploying software on general-purpose devices instead of specialized devices \cite{herrera2016resource}, \cite{7926921}. NFV can tackle the differences and isolation among multiple resource nodes in different networks to realize interconnections and resource sharing for different tasks. Based on NFV, the implementation of resource allocation can be deemed as service function chains (SFCs) deployment \cite{20233814771268}. In detail, SFC is a sequence of virtual network functions (VNFs) that are executed in a certain order to deliver specific network services \cite{10207691}. Consequently, the SFC-based mechanism provides ideas for the resource management in SAGIN, but the following challenges should be focused.

\begin{itemize}
    \item[$\bullet$] In SAGIN, the key components such as satellites and UAVs are highly dynamic, and the relative motion among different nodes leads to dynamic network topology. Therefore, it is challenging to accurately characterize the multi-layer resources across time, which further aggravates the difficulty to establish the SFC deployment model.
    \item[$\bullet$] With the increasing numbers and types of tasks, it is prone to emerging unexpected demands. The urgent unexpected tasks may bring conflicts in SFC deployment and competitions for resources, which in turn leads to task failures.
    \item[$\bullet$] Due to the heterogeneity of satellites and UAVs in terms of dynamics, coverages, and resource capacity, it is challenging to efficiently allocate such heterogenous resources to satisfy multiple SFC demands.
\end{itemize}

As such, in this paper, we focus on dealing with these challenges. Firstly, due to the highly dynamic characteristics of SAGIN, we propose the reconfigurable time extension graph (RTEG) to depict the multiple resources in SAGIN, by dividing the time horizon into multiple time slots, and the network topology is deemed as quasi-static in each time slot. In addition, according to NFV, each task corresponds to an SFC, and the SFC deployment and dynamic scheduling model is designed based on RTEG. Then, the SFC scheduling problem is formulated to maximize the number of successful deployed SFCs, i.e., tasks. Since the problem is in the form of integer linear programming (ILP), direct solutions are intractable due to the unacceptable time complexity \cite{vazirani2001approximation}. Hence, we transform the problem into a Markov decision process (MDP) and further design the deep reinforcement learning (DRL)-based algorithms. Finally, extensive simulations are conducted to verify the performance of the proposed algorithms.

In summary, the main contributions of this work are summarized as follows:
\begin{itemize}
    \item[$\bullet$] We address the dynamics and heterogeneities of SAGIN by proposing RTEG for resource representation, and propose the detailed model of SFC deployment and scheduling based on RTEG for resource allocation.
    \item[$\bullet$] To solve the formulated problem, we transform it into an MDP, and then propose the DRL-based algorithms, in which the algorithm for the mutual selection of SFCs and nodes is designed, so that SFCs can be effectively scheduled and the resources are utilized efficiently. The complexity of the algorithms is also analyzed.
    \item[$\bullet$] The feasibility and efficiency of the proposed algorithms are evaluated through extensive simulations with corresponding analyses, and satisfactory solutions are efficiently obtained for the SFC scheduling problem. 
\end{itemize}

The rest of this paper unfolds as follows. The related work is described in Section \ref{sec:Related-Work}. The system model is proposed in Section \ref{sec:System-Model}, and the problem formulation is presented in Section \ref{sec:Problem-Formulation}. In Section \ref{sec: DRL algorithm}, the algorithms are designed. Simulation results and corresponding analyses are presented in Section \ref{sec:Simulation-Results}. Finally, conclusions are drawn in Section \ref{sec:Conclusions}.

\section{Related Work\label{sec:Related-Work}}
As for the deployment and scheduling of SFC in terrestrial networks, there exist sufficient researches. For example, the authors in \cite{9637726} proposed a heuristic algorithm with a quantum annealer to solve the VNF scheduling problem in virtual machines. In \cite{9099505}, the authors designed a two-phased algorithm to solve the VNF deployment and flow scheduling problems in distributed data centers. The authors in \cite{9084136} presented a deep Dyna-Q approach to handle the SFC dynamic reconfiguration problem in the Internet of Things (IoT) network. In \cite{9785397}, a game theory-based approach to solve SFC service latency problem at the edge was studied. The authors in \cite{9632419} proposed a dynamic SFC embedding scheme with matching algorithm and DRL in the industrial IoT network. The authors in \cite{9122544} optimized the VNF placement and flow scheduling in mobile core networks. However, these models and mechanisms cannot be directly applied to the multi-layer dynamic SAGIN with high heterogeneity.

There exist some studies of SFC deployment in single layer networks in the air such as the flying ad hoc network (FANET), or satellite network in the space. The authors in \cite{9839646} presented a mathematical framework to solve the VNF placement problem in a FANET. In \cite{10032237}, the authors studied a multiple service delivery problem using SFC in the low earth orbit (LEO) satellite-terrestrial integrated network, and designed an improved response algorithm and an adaptive algorithm to achieve the Nash equilibrium. The authors in \cite{9013347} designed an IoT platform running within software defined network (SDN)/NFV-ready infrastructures, which applied to miniaturized CubeSats. In \cite{9652156}, the authors proposed a new edge-cloud architecture based on machine learning, which studied the UAV resource utilization of SFC embedding. \cite{10238738} leveraged UAV-aided mobile-edge computing and NFV to enhance smart agriculture applications, and it introduced  the decentralized federated learning to optimize NFV function orchestration.The authors in \cite{JIA2024104} proposed an approach based on Asynchronous Advantage Actor-Critic to deploy VNFs with low latency during heterogeneous bandwidth demands. \cite{jia2021vnf} investigated the orchestration of NFV chains in satellite networks, followed by the design of a brand-and-price algorithm combining three methods, and proposed an approximate algorithm based on the beam search. However, these works have not considered the connectivity among multi-layers of SAGIN.

There exist a couple of works related with the SFC or VNF deployment in SAGIN. In \cite{AKYILDIZ2019134}, a novel cyber-physical system spanning ground, air, and space was introduced, which was supported by SDN and NFV techniques. The authors in \cite{9951143} used the federation learning algorithm to figure out the SFC embedding problem in SAGIN, and reconfigured SFC to reduce the service blocking rate. In \cite{9062531}, the authors studied a reconfigurable service provisioning framework and proposed a heuristic greedy algorithm to solve the SFC planning problem in SAGIN. In \cite{10207691}, an iterative alternating optimization algorithm by the convex approximation is used to deal with the SFC deployment and scheduling in SAGIN from the perspective of network operators, so as to maximize the network profit. The authors in \cite{9351537} investigated online dynamic VNF mapping and scheduling in SAGIN, and proposed two Tabu search-based algorithms to obtain suboptimal solutions. In \cite{9749937}, the authors constructed a service model by dividing the network slices and proposed an SFC mapping method based on delay prediction. However, the dynamic topology of SAGIN across time have not been well considered in these works, which is a significant issue and cannot be neglected.

As analyzed above, the SFC deployment and scheduling issues are well studied in the terrestrial network, single aerial UAV network, or single satellite network. However, as far as the authors' knowledge, the researches on SFC scheduling problem in SAGIN is not comprehensive, and most designed algorithms are heuristic. These algorithms can not be well adapted to large-scale networks and complete large-scale tasks. Hence, in this paper, we take into account the connectivity and dynamic of the multi-layer SAGIN, propose the corresponding deployment and scheduling model, and design the DRL-based algorithms to cope with network structures of different scales and diverse numbers of task requirements.

\section{System Model\label{sec:System-Model}}

\begin{table}[!t]
    \renewcommand\arraystretch{1.3}
	\begin{center}
		\caption{KEY NOTATIONS} \label{key notations}
		\begin{tabular}{|p{2cm}|p{6cm}|}
			\hline
			Symbol& Description \\
			\hline
            \hline
			$\mathcal{G} =(\mathcal{N},\mathcal{L})$ & SAGIN model composed of nodes $\mathcal{N}$ and links $\mathcal{L}$.\\
            \hline
			$\mathcal{T}$, $t$, $T$, $\tau$ & Total number of time slots, order number of time slots, set of order number, and time slot length.  \\
            \hline
            $\mathcal{F}_k$, $\mathcal{K}$, $k$ & SFC of the $k$-th task, total number of tasks, and the order number of tasks. \\
            \hline
            $n_i^t$, $f_k^m$ & The $i$-th node in time slot $t$, and the $m$-th VNF of SFC $\mathcal{F}_k$. \\
			\hline
            $x_{n_i^t,f^m_k} $ & Binary variable indicating whether VNF $f^m_k$ of SFC $\mathcal{F}_k$ is deployed on node $n_i^t$.\\
            \hline
            $y_{n_i^t}^k $ & Binary variable indicating whether SFC $\mathcal{F}_k$ passes through node $n_i^t$.\\
            \hline
            $I_k$ & Binary variable indicating whether SFC $\mathcal{F}_k$ are deployed successfully.\\ 
            \hline
            $z^k_{(n^t_i,n^t_j)}$ & Binary variable indicating whether SFC $\mathcal{F}_k$ is deployed on link $(n^t_i,n^t_j)$.  \\
            \hline
            $\varrho  ^k_{(n^t_i,n ^{t+1}_i )} $ & Binary variable indicating whether SFC $\mathcal{F}_k$ storages at node $n^t_i$ from time slot $t$ to $t$+1.\\
            \hline
            $\delta_k$ & Data amount of SFC $\mathcal{F}_k$.\\
            \hline
            $\sigma _{f_k^m}$ & Computing resource required by VNF $f_k^m$.\\
            \hline
            $e^c$ & Energy consumption of per unit of computing resource.\\
            \hline
		\end{tabular}
	\end{center}
\end{table}

In this section, the models of network, SFC dynamic scheduling, channel, as well as energy cost are elaborated. Key notations are listed in Table \ref{key notations}.

\subsection{Network Model}

\begin{figure}[!t]
    \centerline{\includegraphics[width=8.3cm]{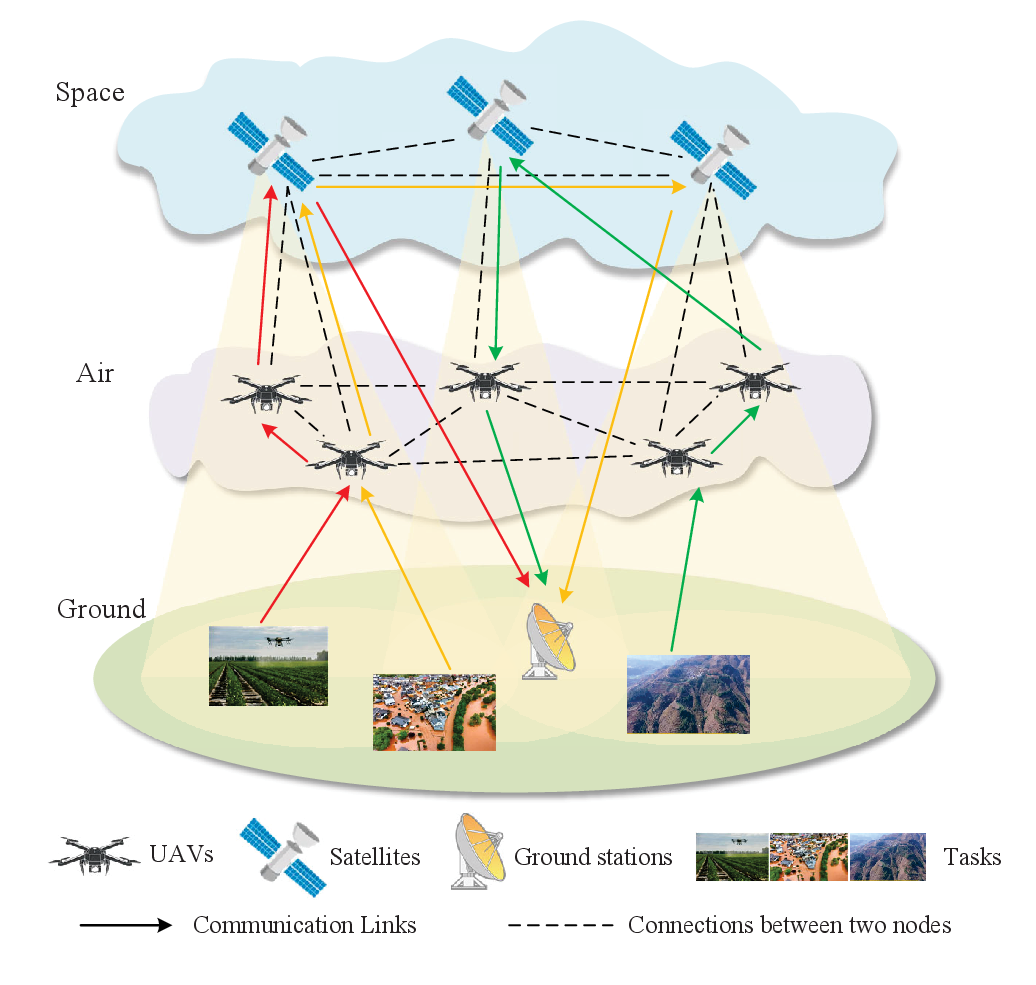}}
   \caption{Scenario of SAGIN.\label{fig1}}
\end{figure}

 The SAGIN scenario includes ground nodes, UAVs in the air and LEO\footnote{Since only LEO satellites are considered in the model of this work, we use the term "satellite" for the LEO satellite in the rest of the paper.} satellites in the space. We conduct an SFC deployment model in SAGIN, as shown in Fig. \ref{fig1}. In detail, it is characterized as $\mathcal{G} =(\mathcal{N},\mathcal{L})$, where $\mathcal{N} =\mathcal{N} _g \cup \mathcal{N} _u \cup \mathcal{N} _s$ represents three types of nodes, $n_i^t \in \mathcal{N}$. Denote $\mathcal{L} =\mathcal{L}_{gu} \cup \mathcal{L}_{sg} \cup \mathcal{L}_{ug}\cup \mathcal{L}_{uu} \cup \mathcal{L}_{us} \cup \mathcal{L}_{ss}  \cup \mathcal{L}_{t}$ as all links between two nodes, i.e., ground-to-UAV (G2U), satellite-to-ground (S2G), UAV-to-ground (U2G), UAV-to-UAV (U2U), UAV-to-satellite (U2S), and satellite-to-satellite (S2S), $(n_i^t,n_j^{t'}) \in \mathcal{L}$. To represent the multiple and heterogenous resources in SAGIN, we design the RTEG, which divides a time horizon into a set of $T$ time slots, $t\in T$. $\mathcal{T}$ is the total number of time slots. Each time slot $t$ has the same length $\tau$, which is sufficiently short so that the link is quasi-static in the same time slot. The same node can be regarded as various nodes in diverse time slots, with different states and resource conditions. In addition, if the data in node $n_i^t$ cannot be transmitted in current time slot $t$, it is stored in the node to next time slot $t$$+1$. Therefore, link $\mathcal{L}_{t}=\{(n_i^t,n_i^{t+1}) | n_i^t \in \mathcal{N} _u \cup \mathcal{N} _s,t\in T\}$ is introduced to indicate the storage of node $n_i$ from time slot $t$ to $t$$+1$.

\begin{figure*}[!t]
    \centerline{\includegraphics[width=18cm]{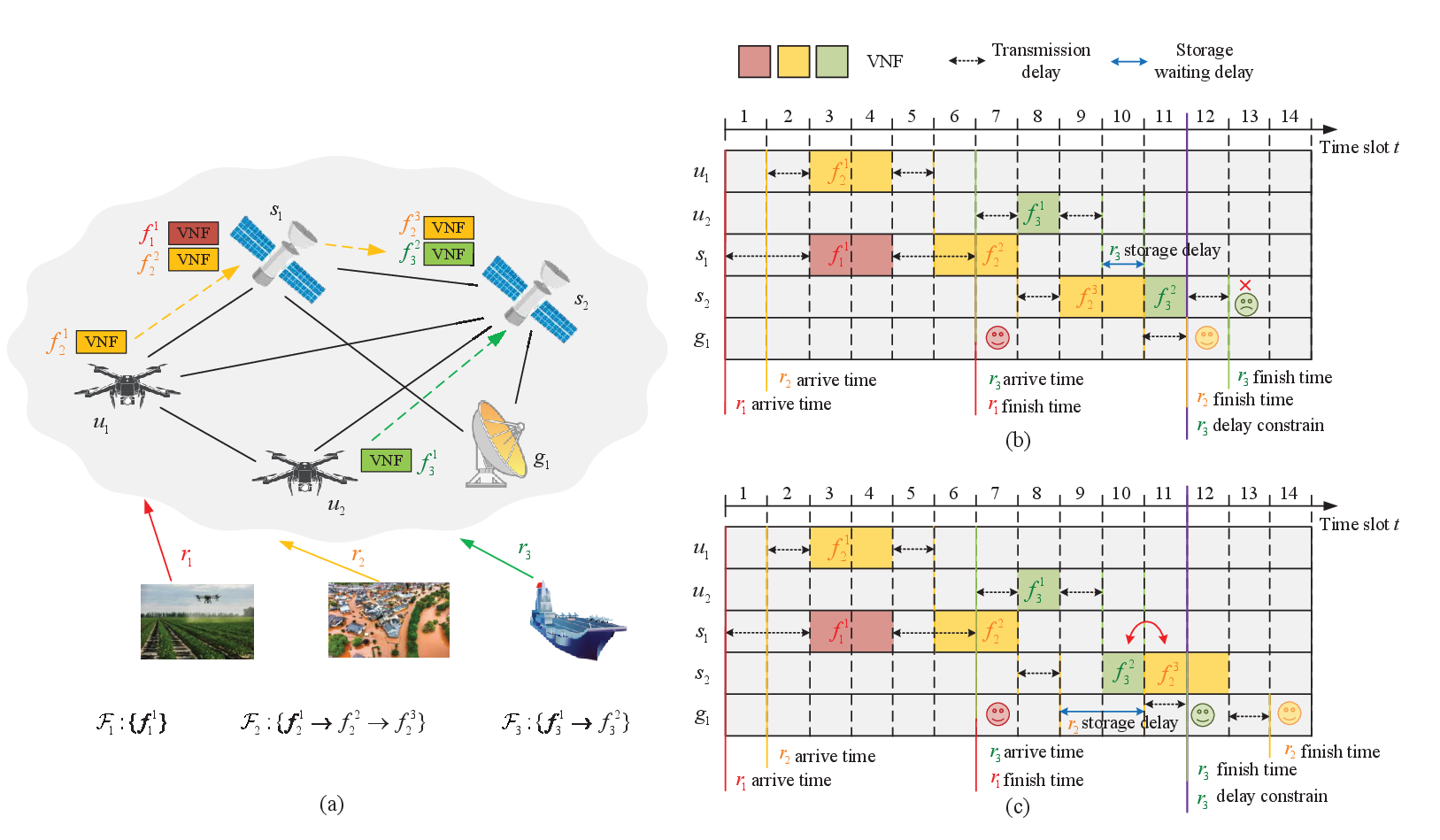}}
   \caption{SFC dynamic scheduling in SAGIN based on RTEG. (a) SFC deployment in SAGIN. (b) SFC offline deployment corresponding to (a). (c) SFC online scheduling.\label{fig2}}
\end{figure*}

\subsection{SFC Dynamic Scheduling Model}

Based on RTEG, we build the SFC scheduling model in SAGIN, as shown in Fig. \ref{fig2}. In detail, there exist three tasks $\{r_1,r_2,r_3\}$ in Fig. \ref{fig2}(a). VNFs are deployed on UAVs or satellites to construct SFCs, where $\mathcal{F}_1:\{f^1_1\}$, $\mathcal{F}_2:\{f^1_2 \rightarrow f^2_2 \rightarrow f^3_2\}$ and $\mathcal{F}_3:\{f^1_3 \rightarrow f^2_3\}$. The SFC is expressed as $\mathcal{F}_k:\{f^1_k \rightarrow f^2_k \rightarrow \cdots \rightarrow f^{l_k}_k\}$, where $f^m_k$ represents the $m$-th VNF in the SFC of the $k$-th task, and $l_k$ is the number of VNFs in the SFC, $f^m_k\in \mathcal{F} _k$. $\mathcal{K}$ represents the number of SFCs, $k \in \mathcal{K}$. It is worth noting that VNFs should be executed in the time order to keep SFC sequence.

Correspondingly, the initial deployment of the three SFCs is shown in Fig. \ref{fig2}(b). Due to the resource restrictions, each node (satellite or UAV) in a time slot can only accommodate limited VNFs. For instance, at time slot 10, VNF $f^2_3$ reaches satellite $s_2$ from UAV $u_2$, but at the same time, there exists another VNF $f^3_2$ being processed on satellite $s_2$. Hence, VNF $f^2_3$ can only being stored on satellite $s_2$, generating the waiting delay until time slot 11. When VNF $f^3_2$ processing is completed and satellite $s_2$ is idle, VNF $f^2_3$ processing can be carried out. However, it makes $f^2_3$ exceed the delay requirement, i.e., SFC deployment and processing is not completed due to latency restrictions.

Hence, we present another scheme for SFC scheduling in Fig. \ref{fig2}(c). When SFC $\mathcal{F}_2$ transmits from satellite $s_1$ to $s_2$ at time slot 9, satellite $s_2$ does not process VNF $f^3_2$ immediately, but stores VNF $f^3_2$. When SFC $\mathcal{F}_3$ is transmitted from UAV $u_2$ to satellite $s_2$ at time slot 10, VNF $f^2_3$ is handled firstly. After completing VNF $f^2_3$, VNF $f^3_2$ is processed. Thus, both SFC $\mathcal{F}_2$ and SFC $\mathcal{F}_3$ satisfy the delay requirements.

It is worth noting that all tasks can be transmitted to nodes within effective communication distance and with sufficient resources. All nodes can effectively handle tasks within the scope of node communication, without node failure and link disconnection. In this case, a comprehensive SFC scheduling scheme is executed for all tasks.

\subsection{Channel Model}

In Fig. \ref{fig1}, there exist six types of channels, including
G2U, U2U, U2S, S2S, S2G and U2G, where the channel types of U2S and S2S are all related to line-of-sight (LoS) communication.

\subsubsection{\textcolor{black}{Channel Model of G2U and U2G}}

Since the height of UAVs is much higher than the ground, we consider that the communication between ground stations and UAVs is LoS, ignoring the small-scale fading and shadow \cite{7572068},\cite{9714482}. Hence, the channel power gain between UAV $n_u^t$ and ground station $n_g$ at time slot $t$ is
\begin{equation}
    G_{gu}^t = \frac{G _0 }{(d^t_{gu})^2}=\frac{G_0}{(a^t_u-a_g)^2+(b^t_u-b_g)^2+h^2_u},
\end{equation}
where $G_0$ is the channel gain when the link distance between the ground and UAV $d^t_{gu}=1m$. Wherein, $\{a^t_u,b^t_u\}$ and $\{a^t_g,b^t_g\}$ are the horizon locations of the UAV and the ground station, respectively. Then, the signal-to-noise ratio (SNR) is 
\begin{equation}
    \Psi _{gu} = \frac{P^{tr} G_{gu}^t }{\sigma_0 ^2}=\frac{P^{tr} \iota_0 }{(d^t_{gu})^2}, 
\end{equation}
where $P^{tr}$ includes $P^{tr}_g$ and $ P^{tr}_u$, which indicate the transmission power from the ground and the UAV, respectively. $\sigma_0^2$ denotes the White Gaussian noise power, and $\iota_0=G_0/\sigma_0 ^2$ is the reference SNR \cite{8594571}, \cite{8937793}.

\subsubsection{Channel Model of U2U}

Following \cite{8411465}, the path loss of U2U is calculated as
\begin{equation}
    PL _{uu} = 20{\rm log}_{10}(d^t_{uu})+20{\rm log}_{10}(f_{uu})-147.55, 
\end{equation}
where $d^t_{uu}$ indicates the distance between two UAVs, and $f_{uu}$ is the frequency. The SNR of U2U can be expressed as
\begin{equation}
    \Psi _{uu} = \frac{P_{uu} 10^{- \frac{PL _{uu}}{10}} }{\sigma_{uu} ^2}, 
\end{equation}
where $P_{uu}$ and $\sigma_{uu} ^2$ denote the transmission power and noise power between two UAVs, respectively.

\subsubsection{\textcolor{black}{Channel Model Related to Satellites}}

Following \cite{TheFederatedSatellite}, the available data rate of link $(n_i^t,n_j^t) \in \mathcal{L}_{us} \cup \mathcal{L}_{ss} $ is expressed as 
\begin{equation}
    r^s_{(n^t_i,n^t_j)}=\frac{P_{ij}G_{ij}^{tr}G_{ij}^{re}L_{ij}L_l}{(E_b/N_0)_{req}k_BT_sS},
\end{equation}
where $r^s_{(n^t_i,n^t_j)}$ includes $r^{us}_{(n^t_i,n^t_j)}$ and $r^{ss}_{(n^t_i,n^t_j)}$, which denote the available data rate of U2S and S2S, respectively. $P_{ij}$ is the transmission power from UAV or satellite $n^t_i$ to satellite $n^t_j$. $G_{ij}^{tr}$ and $G_{ij}^{re}$ represent the transmitting and receiving antenna gains, respectively.  $L_l$ is the total line loss. $(E_b/N_0)_{req}$ denotes the required ratio of the received energy-per-bit to noise density. $k_B$ is the Boltzmann constant. $T_s$ indicates the noise temperature of the total system, and $S$ denotes the maximum slant range. Moreover, $L_{ij}$ is related to the free space loss, i.e.,
$L_{ij}=\left( \frac{c}{4\pi S^t_{ij}f^{cen}}\right)^2$,
where $S^t_{ij}$ is the maximum slant range in time slot $t$. $f^{cen}_{ij}$ refers to the centering frequency.

The S2G channel is affected by atmospheric precipitation, so that the meteorological satellites are used to predict the S2G channel state\cite{9583591}\cite{8353906}. Accordingly, the SNR of S2G is 

\begin{equation}
    \Psi_{sg}=\frac{P_{sg}G_{sg}^{tr}G_{sg}^{re}L_{ij}L_{r}}{N_{0}B_{sg}},
\end{equation}
where $P_{sg}$ represents the transmission power. $G_{sg}^{tr}$ denotes the transmitter antenna gain of satellites, and $G_{sg}^{re}$ indicates the receiver antenna gain of the ground station. $B_{sg}$ represents the bandwidth of S2G. $L_r$ is the rain attenuation and can be acquired from ITU-R P.618-12, i.e., $L_r=L_e \cdot \gamma_R^t$, where $L_e$ is the slant-path length \cite{ITU-R-P618-12}, and $\gamma_R^t$ denotes the attenuation per kilometer in time slot $t$.

According to Shannon formula, the maximum data rate of G2U, U2U, U2G and S2G can be calculated as  
\begin{equation}
    r_{(n^t_i,n^t_j)}=\!B\mathrm{log}{}_{2}(1+\Psi), \forall (n^t_i,n^t_j) \in \mathcal{L} \setminus \{\mathcal{L}_{us} \cup \mathcal{L}_{ss} \cup \mathcal{L}_{t} \},
\end{equation}
where $r_{(n^t_i,n^t_j)}=r^{gu}_{(n^t_i,n^t_j)}\cup r^{uu}_{(n^t_i,n^t_j)}\cup r^{ug}_{(n^t_i,n^t_j)}\cup r^{sg}_{(n^t_i,n^t_j)}$, and $B$ = $B_{gu}\cup B_{uu}\cup B_{ug}\cup B_{sg}$ indicates the bandwidth of different channels. $\Psi=\Psi_{gu}\cup \Psi_{uu}\cup \Psi_{ug}\cup \Psi_{sg}$ denotes the SNR of G2U, U2U, U2G and S2G, respectively.

\subsection{\textcolor{black}{Energy Cost Model}}

\subsubsection{Energy Cost of UAVs}

UAVs primarily consume energy during hovering, moving, and communication\cite{9865119}. The moving power is expressed as 
\begin{equation}
    P^{\rm M}_{n^t_i}=\frac{v_{n^t_i}}{v^{\rm max}_{n^t_i}}(P^{\rm max}_{n^t_i}-P^{\rm H}_{n^t_i}), \forall n^t_i\in\mathcal{N}_{u},t\in T,
\end{equation}
in which $v_{n^t_i}$ denotes the moving speed of UAV $n^t_i$, and $v^{\rm max} _{n^t_i}$ is the maximum speed. $P^{\rm max}_{n^t_i}$ represents the power at the UAV's maximum speed, and $P^{\rm H}_{n^t_i}$ indicates the hovering power, i.e.,
\begin{equation}
    P ^{\rm H}_{n^t_i}= \Theta \sqrt{\frac{(M_{n^t_i}) ^3}{\mu ^2_{n^t_i} \nu _{n^t_i}}} ,\forall n^t_i\in\mathcal{N}_{u},t\in T,
\end{equation}
where $\Theta =\sqrt{{g^3}/{(2\pi\vartheta) }}$ represents the environmental parameter. $g$ is the earth gravity acceleration, and $\vartheta $ indicates the air density. $M_{n^t_i}$ denotes the mass of UAV $n^t_i$, $\mu_{n^t_i} $ is the radius, and $\nu_{n^t_i} $ represents the number of propellers in UAV $n^t_i$. Therefore, the path energy cost of UAV $n^t_i$ is calculated as 
\begin{equation}
    E_{n^t_i,u}^{\rm P}=P^{\rm M}_{n^t_i} \frac{\Vert p _{n^t_i}-p _{n^{t+1}_i}\Vert _2}{v_{n^t_i}}+P ^{\rm H}_{n^t_i}\tau, \forall n^t_i\in\mathcal{N}_{u},t\in T,
\end{equation}
where $p_{n^t}$ denotes the geographical position of UAV $n^t_i$. Besides, the total communication energy cost of UAVs is 
\begin{equation}
    E_{n^t_i,u}^{\rm O}=\underset{k\in\mathcal{K}}{\sum}\underset{n^t_j \in\mathcal{N}}{\sum}\frac{P^{tr}_{n^t_i}z ^k_{(n^t_i,n^t_j)} \delta_k}{r^u_{(n^t_i,n^t_j) }}, \forall n^t_i\in\mathcal{N}_{u},t\in T,
\end{equation}
in which $P^{tr}_{n^t_i}$ represents the transmitted power of UAV $n^t_i$. Binary variable $z ^k_{(n^t_i,n^t_j)}=1$ denotes SFC $\mathcal{F}_k$ is deployed on link $(n^t_i,n^t_j)$; otherwise $z ^k_{(n^t_i,n^t_j)}=0$. $\delta_k$ indicates the data amount of SFC $\mathcal{F}_k$. Hence, the total energy cost can be expressed as 

\begin{equation}
E_{n^t_i,u}^{total}=E_{n^t_i,u}^{\rm P}+E_{n^t_i,u}^{\rm O},\forall n^t_i\in\mathcal{N}_{u},t\in T. \label{formula:energycost of HAP}
\end{equation}

\subsubsection{\textcolor{black}{Energy Cost of Satellites}}
The energy consumption of satellites is primarily associated with the data transmission and reception. $E_{n^t_i,s}^{re}$ denotes the energy cost of a data receiver, while $E_{n^t_i,s}^{tr}$ indicates the transmitter energy cost of satellites. Therefore, we have: $\forall n^t_i\in\mathcal{N}_{s},t\in T,$
\begin{equation}
    E_{n^t_i,s}^{re}\!=\!\underset{k\in\mathcal{K}}{\sum}\!\left(\underset{n^t_j \in\mathcal{N}_{u}}{\sum}\!\!\frac{P_{us}^{re}z^k_{(n^t_j,n^t_i)} \delta_k}{r^{us}_{(n^t_j,n^t_i) }}\!+\!\!\!\underset{n^t_j\in\mathcal{N}_{s}}{\sum}\!\!\frac{P_{ss}^{re}z^k_{(n^t_j,n^t_i)} \delta_k}{r^{ss}_{(n^t_j,n^t_i) }}\!\right)\!, \label{formula: Energycost LEO}
\end{equation}
and
\begin{equation}
    E_{n^t_i,s}^{tr}\!=\!\underset{k\in\mathcal{K}}{\sum}\!\left(\underset{n^t_j \in\mathcal{N}_{s}}{\sum}\!\!\frac{P_{ss}^{tr}z^k_{(n^t_i,n^t_j)} \delta_k}{r^{ss}_{(n^t_i,n^t_j) }}\!+\!\!\!\underset{n^t_j\in\mathcal{N}_{g}}{\sum}\!\!\frac{P_{sg}^{tr}z^k_{(n^t_i,n^t_j)} \delta_k}{r^{sg}_{(n^t_i,n^t_j) }}\!\right)\!, 
\end{equation}
where $P_{us}^{re}$ and $P_{ss}^{re}$ denote the received power of U2S and S2S, respectively. $P_{ss}^{tr}$ and $P_{sg}^{tr}$ represent the transmitted power of S2S and S2G, respectively. Hence, the total energy consumption of satellites is 
\begin{equation}
E_{n^t_i,s}^{total}=E_{n^t_i,s}^{re}+E_{n^t_i,s}^{tr}+E_{n^t_i,s}^{op}, \forall n^t_i\in\mathcal{N}_{s},t\in T,
\end{equation}
in which $E_{n^t_i,s}^{op}$ indicates the general operation energy cost.

\section{Problem Formulation\label{sec:Problem-Formulation}}

\subsection{Constraints}
\subsubsection{Deployment Constraints}
We introduce binary variable $x_{n_i^t,f^m_k}=1 $ to indicate VNF $f^m_k$ in SFC $\mathcal{F}_k$ is deployed on node $n_i^t$; otherwise $x_{n_i^t,f^m_k}=0 $. Each VNF can only be deployed on one node, i.e.,
\begin{equation}
    \underset{n_i^t \in \mathcal{N}_u\cup\mathcal{N}_s}{\sum} \ x_{n_i^t,f^m_k}=1,\ \forall  f^m_k \in \mathcal{F} _k.\label{cons:x01}
\end{equation}
Correspondingly, a finite number of different VNFs can be deployed on one node at the same time. Besides, VNF $f^m_k$ can be deployed only when SFC $\mathcal{F}_k$ passes through node $n_i^t$. Consequently, we have 
\begin{equation}
    x_{n_i^t,f^m_k}\leq y_{n_i^t}^k,\ \forall  f^m_k\in \mathcal{F} _k,n_i^t\in \mathcal{N},\label{cons:y01}
\end{equation}
where binary variable $y_{n_i^t}^k =1 $ denotes that SFC $\mathcal{F}_k$ passes through node $n_i^t$; otherwise $y_{n_i^t} =0 $.

In addition, we introduce binary variable $I_k$ to denote whether all VNFs of SFC $\mathcal{F}_k$ are deployed successfully, i.e.,
\begin{equation}
    I_k=
    \left\{
        \begin{aligned}
           1, \;\;& \text{if} \;\underset{n_i^t \in \mathcal{N}_u\cup\mathcal{N}_s}{\sum}\underset{f^m_k \in \mathcal{F}_k}{\sum} x_{n_i^t,f^m_k}=l_k, \;\\
           0, \;\;& \text{otherwise},
        \end{aligned}
    \right.\;\forall k \in \mathcal{K}.
\end{equation}

\subsubsection{Sequence Constraints for SFC}

The VNFs of SFC $\mathcal{F}_k: \{f^1_k \rightarrow f^2_k \rightarrow \cdots \rightarrow f^{l_k}_k\}$ must be deployed sequentially\cite{9075271}:
\begin{equation}
    t_{f^{m+1}_k}-t_{f^m_k} \geq \underset{t \in T}{\sum}\underset{n_i^t\in \mathcal{N}_u\cup\mathcal{N}_s}{\sum}\!\! \frac{ x_{n_i^t,f^m_k} \sigma _{f_k^m}}{\varphi _{n^t_i}} , \forall  f^m_k\in \mathcal{F} _k,\label{cons:tt01}
\end{equation}
where $t_{f^m_k}$ denotes the time slot when VNF $f^m_k$ starts processing. Let $\sigma _{f_k^m}$ (bit) represent the computing resource required by VNF $f_k^m$, and $\varphi _{n^t_i}$ (bit/s) indicate the computation ability of node $n^t_i$.

\subsubsection{Flow Constraints}
The SFC deployment must satisfy the flow conservation constraints:

\begin{subnumcases} {\label{flowcons}}
    \sum_{(n^t_o,n^t_j)\in \mathcal{L}\setminus \mathcal{L}_t}\!\!\!\! z^k_{(n^t_o,n^t_j)}=1,\forall  k \in \mathcal{K}, n^t_o \in \mathcal{N}, t \in T, \label{cons:flowconsOD1}\\ 
    \sum_{(n^t_i,n^t_j)\in \mathcal{L}\setminus \mathcal{L}_t}^{} \!\!\!\!z^k_{(n^t_i,n^t_j)}+\!\!\!\sum_{(n^{t-1}_j,n^t_j)\in \mathcal{L}_{t-1}}^{} \!\!\!\!z^k_{(n^{t-1}_j,n^t_j)}\!=\!\!\!\! \nonumber\\\
    \sum_{(n^t_j,n^t_i)\in \mathcal{L}\setminus \mathcal{L}_t}^{} \!\!\!\!z^k_{(n^t_j,n^t_i)}+\!\!\sum_{(n^t_j,n^{t+1}_j)\in \mathcal{L}_t}^{} \!\!\!z^k_{(n^t_j,n^{t+1}_j)}, \nonumber\\
    \;\;\;\;\;\;\;\;\;\;\;\;\;\;\;\;\;\;\;\;\;\forall  k\in \mathcal{K}, n^t_j \in \mathcal{N}, t \in \{2,...,\mathcal{T}\!\!\!-1\}, \label{cons:flowconserS}\\
    \sum_{(n^t_i,n^t_d)\in \mathcal{L}\setminus \mathcal{L}_t} z^k_{(n^t_i,n^t_d)}=1,\forall  k \in \mathcal{K}, n^t_d \in \mathcal{N}, t \in T, \label{cons:flowconsOD2}
\end{subnumcases}
where $n^t_o$ and $n^t_d$ denote the original node and the destination node of the SFC deployment, respectively. Eqs. (\ref{cons:flowconsOD1}), (\ref{cons:flowconserS}) and (\ref{cons:flowconsOD2}) represent the flow constraints at the start, intermediate, and end nodes, respectively. Binary variable $z ^k_{(n^t_i,n^t_j)}=1$ denotes SFC $\mathcal{F}_k$ is deployed on link $(n^t_i,n^t_j)$; otherwise $z ^k_{(n^t_i,n^t_j)}=0$. The binary variable $z^k_{(n^{t-1}_j,n^t_j)}=0$ if $t=1$, and $z^k_{(n^t_j,n^{t+1}_i)}=0$ if $t=\mathcal{T}$.

For SFC $\mathcal{F}_k$, only one of three situations can occur in time slot $t$, including deployment on a node, transmission on a link, or storage on a node. Therefore, we have: $\forall k\in \mathcal{K},t \in T,$
\begin{equation}
    \underset{n_i^t \in \mathcal{N}_u\cup\mathcal{N}_s}{\sum} \!\!\! x_{n_i^t,f^m_k} +\!\!\! \sum_{(n^t_i,n^t_j)\in \mathcal{L}\setminus \mathcal{L}_t}^{} \!\!\!z^k_{(n^t_i,n^t_j)} + \!\!\!\sum_{(n^t_i,n ^{t+1}_i)\in \mathcal{L}_t}^{}  \!\!\!\!\!\varrho  _{k,{(n^t_i,n ^{t+1}_i )}} = 1,  \label{cons:xyzv}
\end{equation} 
where $\varrho  ^k_{(n^t_i,n ^{t+1}_i )}=1 $ indicates SFC $\mathcal{F}_k$ is stored on node $n^t_i$ from time slot $t$ to $t$+1; otherwise $\varrho  ^k_{(n^t_i,n ^{t+1}_i )}=0 $.

\subsubsection{Resource Constraints}

The total computing resource for SFCs on the deployed nodes cannot exceed the computation capacity, i.e.,
\begin{equation}
    \sum_{k\in \mathcal{K} }^{}  \sum_{f^m_k\in \mathcal{F}_k}^{} x_{n_i^t,f^m_k} \sigma _{f_k^m}\leq C_ {n_i^t} ,\forall n_i^t \in \mathcal{N}_u \cup \mathcal{N}_s, t \in T,\label{cons:comcapacity1}
\end{equation}
where $C_{n_i^t}$ includes $C_{n_i^t}^u$ and $ C_ {n_i^t}^s$, which denote the resource capacity of UAVs and satellites, respectively.

When SFC $\mathcal{F}_k$ waits to be processed on node $n^t_i$, it consumes the storage resources of node $n^t_i$. Hence, the total stored data cannot exceed the storage capacity $U_{n^t_i}$, i.e.,
\begin{equation}
    \sum_{k\in \mathcal{K}}^{}  \varrho  _{k,{(n^t_i,n ^{t+1}_i )}} \delta_k  \leq U_{n^t_i},\forall (n^t_i,n ^{t+1}_i)\in \mathcal{L}, t \in T.\label{cons:storageCapacity}
\end{equation}
SFCs with varying data amount transmit in diverse links, the channel capacity restriction should be satisfied:
\begin{equation}
    \sum_{k\in \mathcal{K}}^{}  z^k_{(n^t_i,n^t_j)} \delta_k\leq r_{(n^t_i,n^t_j)}\tau,\forall (n^t_i,n^t_j)\in \mathcal{L} \setminus \mathcal{L}_t, t\in T.\label{cons:links1}
\end{equation}

Furthermore, the total energy consumption cannot exceed the energy capacity $E^{max}_{n^t_i}$, i.e.,
\begin{equation}
    E^c_{n^t_i}+E^{total}_{n^t_i}\leq E^{max}_{n^t_i} ,\forall {n^t_i} \in \mathcal{N}_u \cup \mathcal{N}_s,t \in T,\label{cons:energytotal}
\end{equation}
where $E^{total}_{n^t_i}$ denotes the transmission energy cost. Besides, the energy cost for computation is expressed as
\begin{equation}
    E^{c}_{n^t_i}=\sum_{k\in \mathcal{K} }^{}  \sum_{f^m_k\in \mathcal{F} _k}^{}  x_{n^t_i,f^m_k} \sigma _{f^m_k} e^c, \forall n^t_i\in\mathcal{N}_u \cup \mathcal{N}_s, t \in T,\label{cons:energycom1}
\end{equation}
where $e^c$ includes $e^c_u $ and $ e^c_s$, which are the energy consumption of per unit computing resource on a UAV and a satellite node, respectively. $E^{c}_{n^t}$ includes $E^{c}_{n^t,u} $ and $ E^{c}_{n^t,s}$, which denote the computing energy cost of UAVs and satellites, respectively.

\subsubsection{Delay Constraints}

The total time cost of SFC $\mathcal{F}_k$ deployment cannot exceed the maximum tolerable delay $D^{max}_k$, i.e.,
\begin{equation}
    t^{f}_k+ t^{tr}_k +\!\!\!\underset{(n^t_i,n ^{t+1}_i)\in \mathcal{L}_t}{\sum}\!\!\!\varrho  _{k,{(n^t_i,n ^{t+1}_i )}}  \leq D^{max}_k, \forall k\in \mathcal{K},\label{cons:timedelay}
\end{equation}
where 
\begin{equation}
    t^f_k = \underset{n^t_i \in \mathcal{N}_u\cup\mathcal{N}_s}{\sum}\underset{f^m_k\in \mathcal{F}_k}{\sum}  \frac{ x_{n_i^t,f^m_k} \sigma _{f_k^m}}{\varphi _{n^t_i}}  , \forall k\in \mathcal{K}, \label{cons:processdelay}
\end{equation}
which denotes the VNF processing delay for SFC $\mathcal{F}_k$, and $t^{tr}_k$ indicates the transmission delay for SFC $\mathcal{F}_k$:
\begin{equation}
    t^{tr}_k= \underset{(n^t_i,n^t_j)\in \mathcal{L}\setminus \mathcal{L}_t}{\sum}\!  \frac{z^k_{(n^t_i,n^t_j)}\delta_k}{r_{(n^t_i,n^t_j)}}  , \forall k\in \mathcal{K}. \label{cons:transdelay}
\end{equation}

\subsection{Optimization Objective}
The objective is to maximize the number of successful deployed SFCs, i.e.,

\begin{equation}{\label{optimal}}
    \begin{aligned}
    \mathscr{P}0:\;&\underset{\boldsymbol{X},\boldsymbol{Y},\boldsymbol{Z},\boldsymbol{V},\boldsymbol{I}   }{\textrm{max}}\; \sum_{k \in \mathcal{K}}I_k \\
    &\begin{array}{r@{}l@{\quad}l}
        \textrm{s.t.} \;\;&(\ref{cons:x01}), (\ref{cons:y01}),(\ref{cons:tt01})-(\ref{cons:energytotal}), (\ref{cons:timedelay}), \\
             &x_{n^t_i,f^m_k}, y_{n^t_i}^k, z^k_{(n^t_i,n^t_j)}, \varrho ^k_{(n^t_i,n ^{t+1}_i )},I_k\in \{0,1\},\\
        \end{array}
    \end{aligned}
\end{equation}
where $\boldsymbol{X}=\{x_{n^t_i,f^m_k},\forall k\in \mathcal{K}, n^t_i \in \mathcal{N}_u \cup \mathcal{N}_s,t\in T\} $, $\boldsymbol{Y}=\{y^k_{n^t_i},\forall k\in \mathcal{K}, n^t_i \in \mathcal{N},t\in T\}$, $\boldsymbol{Z}=\{z^k_{(n^t_i,n^t_j)},\forall k\in \mathcal{K}, (n^t_i,n^t_j)\in\mathcal{L}\setminus \mathcal{L}_t ,t\in T\}$, $\boldsymbol{V} =\{\varrho ^k _{(n^t_i,n ^{t+1}_i )},\forall k\in \mathcal{K}, (n^t_i,n ^{t+1}_i )\in \mathcal{L}_t,t\in T\}$, and $\boldsymbol{I}=\{I_k,\forall k \in \mathcal{K}\} $. 
It is noted that $\mathscr{P}0$ is an ILP problem, which is difficult to solve within limited time complexity \cite{wolsey1999integer}. Hence, in the following section, we design efficient algorithms based on DRL.

\section{algorithm design\label{sec: DRL algorithm}}
\subsection{\textcolor{black}{MDP Transformation}}
As the above discussions, the original problem $\mathscr{P}0$ is intractable to directly solve. Hence, to cope with the dynamically changing and complex problem, we propose an algorithm based on DRL. Besides, the mutual selection of SFCs with nodes and links in SAGIN (MSSNL-SAGIN) is included, and so the algorithm is termed as DRL-MSSNL-SAGIN. Firstly, we consider transforming the SFC scheduling problem as an MDP, which consists of five tuples $\langle \mathcal{S},\mathcal{A},\mathcal{P},\mathcal{R}, \gamma  \rangle $, where $\mathcal{S}$ is the set of system states, $\mathcal{A}$ denotes the action set, $\mathcal{P}$ represents the state transition probability, $\mathcal{R}$ denotes the reward function, and $\gamma $ indicates the discount factor. The tuples are detailed as follows.

\subsubsection{\textcolor{black}{System State}}
We capture the system state at the beginning of each time slot $t$. State $s_k^t$ is divided into two parts: one is related to SFC $\mathcal{F}_k$, and another is related to nodes in SAGIN, i.e.,
\begin{equation}
    s_k^t=\{\boldsymbol{F}_k^t,\boldsymbol{N}^t\},
\end{equation}
where $\boldsymbol{F}_k^t=\{k, v_k^t, \mathcal{C}_k^t\}$ includes the index $k$ of SFC $\mathcal{F}_k$, the state $v_k^t$ of the VNF being processed in SFC $\mathcal{F}_k$, and the node $\mathcal{C}_k^t$ selected by SFC $\mathcal{F}_k$ in the previous time slot. Besides, $\boldsymbol{N}^t=\{\eta _1^t,\eta _2^t,...,\eta_\mathcal{I}^t\}$ represents the resource occupancy of all nodes, where $\mathcal{I}$ denotes the total number of nodes. $\eta_i^t$ indicates the amount of resources already occupied on node $n_i^t$. Moreover, state $v_k^t$ can be further expressed as 
\begin{eqnarray}
v_k^t=
    \begin{cases}
    0, & \mbox{if VNF} \; f_k \; \mbox{is being transmitted on} \;\mathcal{L} \;\mbox{in} \;t, \\
    1, & \mbox{if VNF} \; f_k \; \mbox{is being processed on} \;\mathcal{N} \;\mbox{in} \;t,\\
    2, & \mbox{if VNF} \; f_k \; \mbox{is stored and waiting in} \;t.
    \end{cases}
\end{eqnarray}

\subsubsection{\textcolor{black}{Action}}
Each SFC needs to select an effective node in a time slot if it wants to be successfully deployed and processed on the node. Therefore, action $a_k^t$ is set as the node selected by SFC $\mathcal{F}_k$ in current time slot $t$. Besides, the whole action set contains all SFCs, i.e., $A^t=\{a_1^t,a_2^t,...,a_\mathcal{K} ^t\}$, where $\mathcal{K} $ is the total number of SFCs.

\subsubsection{\textcolor{black}{State Transition}}
By selecting different actions, the state of SFC changes accordingly. Firstly, the pending VNF state $v_k^{*t+1}$ of the next time slot is obtained, and then the determined VNF state $v_k^{t+1}$ is acquired through the selection of nodes. The pending VNF state $v_k^{*t+1}$ is defined as follows:
\begin{eqnarray}
    v_i^{*t+1}=
        \begin{cases}
        0, & \mbox{if} \; v_k^t=0, \;t_k^c(t) > 1, \\
        0, & \mbox{if} \; v_k^t=0, \;t_k^c(t) <= 1, \;a_k^t \neq \mathcal{C}_k^t, \\
        0, & \mbox{if} \; v_k^t=1 \; \mbox{or} \; 2, \;a_k^t \neq \mathcal{C}_k^t, \\
        1, & \mbox{if} \; v_k^t=1, \;t_k^p(t) > 1, \;a_k^t = \mathcal{C}_k^t,\\
        2, & \mbox{if} \; v_k^t=0, \;t_k^c(t) <= 1, \;a_k^t = \mathcal{C}_k^t,\\
        2, & \mbox{if} \; v_k^t=1, \;t_k^p(t) <= 1, \;a_k^t = \mathcal{C}_k^t,\\
        2, & \mbox{if} \; v_k^t=2, \;a_k^t = \mathcal{C}_k^t,
        \end{cases}
    \end{eqnarray}
where $t_k^c(t)$ denotes the transmission time over the channel, which is  consumed by SFC $\mathcal{F}_k$ after selecting a certain action. $t_k^p(t)$ indicates the remaining processing time for VNF $f_k$ currently being processed of SFC $\mathcal{F}_k$.

\subsubsection{\textcolor{black}{Reward}}
Since the optimization objective is to deploy as many SFCs as possible within limited time, SFCs need to be scheduled optimally and efficiently. For each SFC, minimizing the ineffective time consumption, such as waiting time on the busy node, can help improve the optimization objective. Therefore, when an SFC takes an action, the immediate reward is set as
\begin{equation}
    R_k^t=c_0-c_1*t_k^c(t)-c_2*t_k^w(t), \label{cons:reward}
\end{equation}
where $t_k^w(t)$ represents the waiting time at the node selected by SFC $\mathcal{F}_k$. Constants $c_0$, $c_1$ and $c_2$ are weighting coefficients, which are used to adjust the reward value, the weight of transmission time and waiting time consumption, respectively, to ensure that the reward remains within a fixed range.

\begin{figure*}[!t]
    \centerline{\includegraphics[width=16.4cm]{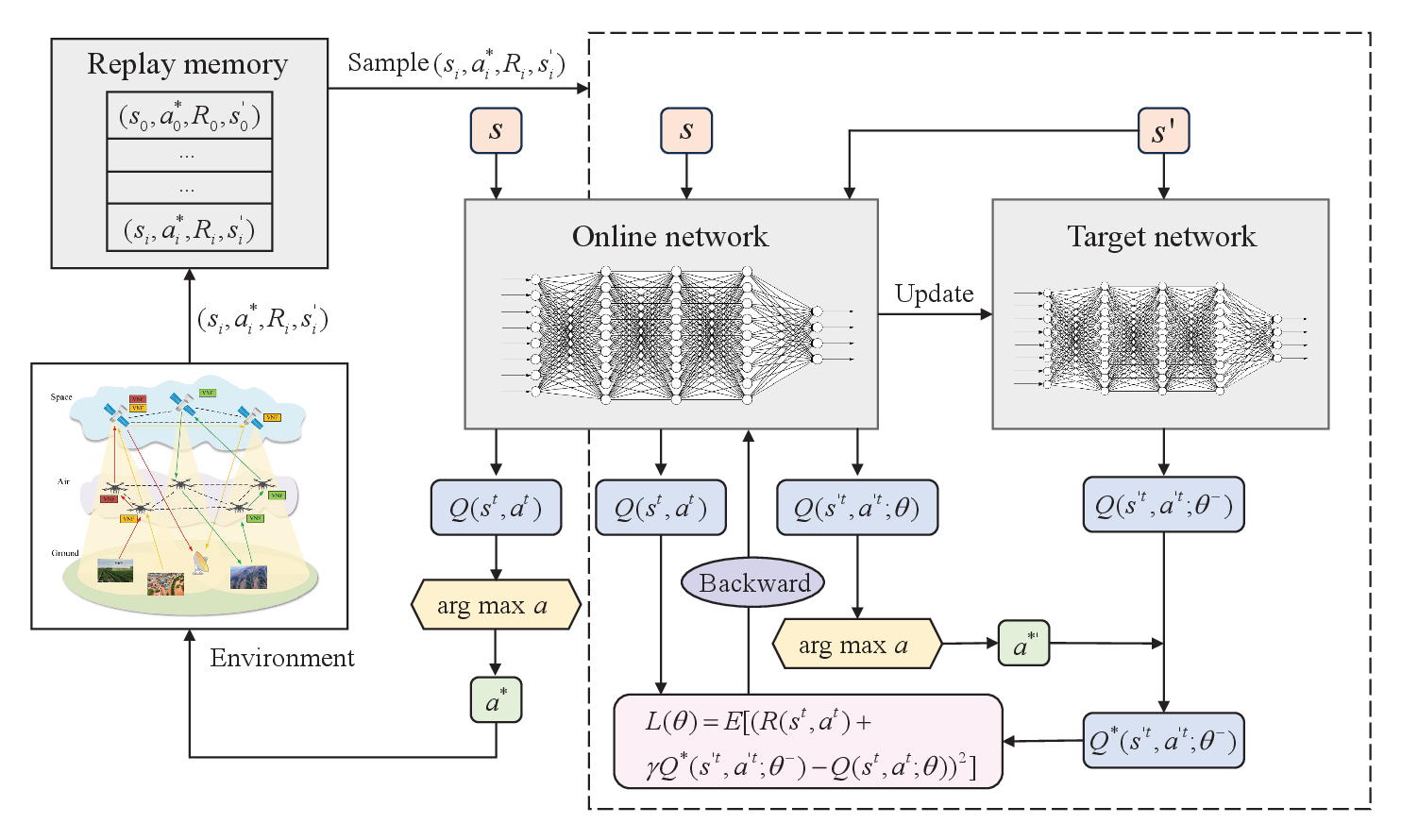}}
   \caption{The training process of DRL-MSSNL-SAGIN algorithm.\label{DDQN strategy}}
\end{figure*}

\subsection{\textcolor{black}{DRL-based Algorithm}}
With the states, actions, state transitions, and rewards, the optimal scheduling policy $\pi^*$ can be obtained by the reinforcement learning (RL) algorithm to maximize rewards over time \cite{filar2012competitive}, \cite{arulkumaran2017deep}. In Q-learning, the optimal policy is obtained by continuous learning. During the learning process, the Q-value table is updated iteratively, i.e., 
\begin{equation}
    \begin{aligned}
        Q(s^t,A^t)\leftarrow &Q(s^t,A^t) +\alpha[R(s^t,A^t) \\ &+\gamma \underset{A^{t+1}}{\textrm{max}}Q(s^{t+1},A^{t+1})-Q(s^t,A^t)] ,\label{cons:Q_update}
    \end{aligned}
\end{equation}
where $\alpha$ denotes the learning rate, and $\gamma  \in [0,1]$ is the discount factor representing the attenuation value of rewards. If $\gamma $ is closer to 1, it is sensitive to future rewards. $Q(s^t,A^t)$ indicates the expected reward for the state-action pair $(s^t,A^t)$, which can express the probability of taking action $A^t$ at state $s^t$. If the Q-value table is able to converge to its optimal Q* after sufficiently large episodes, the optimal policy is obtained as
\begin{equation}
    \pi^*=\textrm{arg} \;\underset{A^t}{\textrm{max}} \;Q^*(s^t,A^t).\label{cons:pi*arg}
\end{equation}

As a basic method of RL, Q-learning performs well in small states and action spaces. However, in this paper, the scale of states and spaces is quite large, so it is intractable to build the Q-value table. Since the introduction of deep neural networks (DNNs) into the framework of Q-learning can deal with the large-scale problem, deep Q network (DQN) is an effective method. However, DQN cannot always guarantee the convergence. The estimated Q-values may fluctuate continuously during training and even fail to converge to the optimal solution. Hence, double deep Q network (DDQN) is a better mechanism for the SFC scheduling problem.

To be specific, the state value is used as the input for DNNs, and all the action values are output. Then, the action with the maximum value is directly selected as the next action. Moreover, there exist two types of networks in DDQN: the online network and target network, and DDQN can stabilize the overall performance by these two networks. In addition, the parameters of the online network are completely copied to the target network at intervals for update. Such delayed updates can ensure the training stability for the Q network. The weights $\theta ^{-}$ of the target network is fixed during the iteration while the weights $\theta $ of the online network are updated. By constantly updating $\theta $, the loss function $L(\theta )$ is minimized, so as to gradually reach the optimal solution. Most steps of DDQN and DQN are similar, but DQN always selects the maximum output value of the target network, while DDQN firstly obtains the action with the maximum output value from the online network, and then acquires the output value of the target network corresponding to this action. Then, the loss function $L(\theta )$ is formulated as
\begin{equation}
    L(\theta)=\mathbb{E}_{s^t,a^t,R(s^t,a^t),s'^t}[(y^{DDQN}-Q(s^t,a^t;\theta ))^2] ,\label{cons:loss}
\end{equation}
and
\begin{equation}
    y^{DDQN} =R(s^t,a^t)+\gamma Q(s'^t,\textrm{arg} \underset{a'^t \in \mathcal{A}^t}{\textrm{max}} Q(s'^t,a'^t;\theta );\theta ^{-}),\label{cons:yDDQN}
\end{equation}
where $\mathbb{E}[\cdot]$ is the expectation operator, $\gamma$ denotes the discount rate, and $\theta ^{-}$ indicates the weights of a target network. The action $a^t$, as mentioned above, can be obtained from the online network $Q(s^t,a^t;\theta)$ with the $\epsilon $-greedy policy by the DNN.


In DDQN, the experience replay memory $\mathcal{D}$ is used to cope with the instability of learning. In detail, after passing through DNN, a new experience $(s^t,a^t,R(s^t,a^t),s'^t)$ is obtained, and the transformed experience is put into $\mathcal{D}$. In this way, small batches of experience from $\mathcal{D}$ are sampled uniformly and randomly to train the neural network. Random sampling reduces the correlation between the training samples, and thus local minima can be avoided during the training process.

The specific training process of DDQN is shown in Fig. \ref{DDQN strategy}. By interacting with the environment, actions and rewards can be obtained, according to the state information. The information are put into the experience replay memory, and a batch is randomly sampled to train the neural network. Then, the best action for the next state is selected by the online network, and the target network is used to evaluate the Q-value of this action to obtain the target Q-value. After that, the loss function is calculated by the Q-value achieved from the online network and the target Q-value. The parameters of the online network are updated by the back-propagation algorithm, and periodically copied to the target network to maintain the stability. It is notable that the training process of DDQN in this paper is an offline mechanism.

Due to the dynamic nature of SAGIN, the distances between nodes change according to time slots, which affect the task deployments of the current time slot. The current SFC deployment scheme may not be optimal in the next time slot, which is a short-sighted local optimal situation. DDQN can cope with the dynamic nature of SAGIN, adjust the deployment situation at any time, and propose different SFC deployment schemes according to different network conditions. The Actor-Critic algorithm or Proximal Policy Optimization algorithm is more suitable for the scene of continuous action space. If these algorithms are applied to the discrete action space, it may cause inefficient calculation. Instead, it increases the computation amount and cost. The action space is simple and discrete. Hence, DDQN is suitable for dealing with discrete action spaces.

Algorithm \ref{DDQN algorithm} provides the detailed DRL-MSSNL-SAGIN procedures for determining the optimal scheduling policy. The movement trajectory of UAVs are clearly known, and the trajectories of satellites are regular. In the whole training process, we consider the position relationships and states of UAV nodes and satellite nodes in different time slots. At the beginning of the training, the relevant values of the online network, target network and replay memory $\mathcal{D}$, as well as the system state are initialized (lines \ref{line1}-\ref{line4}). Each SFC selects an action by the $\epsilon$-greedy policy at the beginning of each time slot in each episode (line \ref{line6}). Then, SFCs obtain the next states based on actions, which includes the VNF states of SFCs, and the node states (lines \ref{line7}-\ref{line8}). The next VNF states of SFCs is updated by Algorithm \ref{RepeatVNFState} (line \ref{line9}). Then, the results on the status of SFC deployment completion are obtained (line \ref{line10}). The neural network model is then optimized by the resulting reward (lines \ref{line11}-\ref{line14}). If all SFCs are successfully deployed, the training round ends; otherwise the loop continues to the maximum value of steps (lines \ref{line15}-\ref{line16}).

\begin{algorithm}[!t]
    \caption{DRL-MSSNL-SAGIN algorithm.\label{DDQN algorithm}}

    \begin{algorithmic}[1]
        \REQUIRE State $s^t$, all nodes $n \in \mathcal{N}_u \cup \mathcal{N}_s$, and set $K$.
        \ENSURE The number of successful deployed SFCs.
        \STATE \textbf{Initialization:} \label{line1}
        \STATE Initialize the replay memory $D$, DDQN network parameter $\theta $, online network $Q(s^t,a^t;\theta)$, and the target network $Q(s^t,a^t;\theta ^-)$ with $\theta ^- =\theta $.
        \FOR {each episode}
            \STATE Initialize state $s^0$ and the node state.\label{line4}
            \WHILE {step $t < T$}
                \STATE Each $\mathcal{F}_k$ chooses $a^t$ using $\epsilon$ -greedy policy.\label{line6}
                \STATE Each $\mathcal{F}_k$ obtains the next pending state based on $a^t$.\label{line7}
                \STATE Update $t_k^c(t)$, $t_k^p(t)$, and the serial number of the currently processing VNF.\label{line8}
                \STATE Obtain $v^{t+1}$ according to Algorithm \ref{RepeatVNFState}.\label{line9}
                \STATE Calculate the reward $R_k^t$ of $\mathcal{F}_k$ according to $(\ref{cons:reward})$.\label{line10}
                \STATE Each $\mathcal{F}_k$ stores transition $(s_k^t,a_k^t,R_k^t,s_k^{t+1})$.\label{line11}
                \STATE Each $\mathcal{F}_k$ samples a batch of transitions from $D$ randomly.\label{line12}
                \STATE Calculate $y^{DDQN}$ according to $(\ref{cons:yDDQN})$.
                \STATE Update parameter $\theta ^{-}$ according to $(\ref{cons:loss})$.\label{line14}
                \IF {all SFCs finish deployments}\label{line15}
                    \STATE Break.\label{line16}
                \ENDIF
            \ENDWHILE
        \ENDFOR
    \end{algorithmic}
\end{algorithm}

\begin{algorithm}[!t]
    \caption{Algorithm for VNF state transition.\label{RepeatVNFState}}

    \begin{algorithmic}[1]
        \REQUIRE VNF state $v^t$, action $a^t$, next pending VNF state $v^{*t}$, and all nodes $n \in \mathcal{N}_u \cup \mathcal{N}_s$.
        \ENSURE Next VNF state $v^{t+1}$.
        \STATE Categorize actions into two types: one corresponding to UAV $\mathcal{N}_u$ and another corresponding to satellite $\mathcal{N}_s$.
        \IF {the node is only selected by one SFC $\mathcal{F}_k$}\label{2line2}
            \STATE VNF state $v_k^{t+1} \leftarrow 1$.  \label{2line3}
        \ELSIF {the node is selected by more than one SFC}
            \IF {the pending state $v_k^{*t+1} = 1$}\label{2line5}
            \STATE The node still accepts the current SFC $\mathcal{F}_k$, and the remaining computational resources of the node is updated.\label{2line6}
            \ENDIF
            \STATE Sort the remaining SFCs that select the node according to the amount of data in an ascending sequence.\label{2line8}
            \STATE Select SFCs according to the remaining computing resources of the node.\label{2line9}
            \STATE Update the next state of SFC $v_k^{t+1} = 1$.
        \ENDIF
    \end{algorithmic}
\end{algorithm}

Moreover, Algorithm \ref{RepeatVNFState} picks SFCs that selecting the same nodes in order to determine the next states of SFCs. In detail, whether the node is selected by more than one SFC is assessed firstly. If there is only one SFC, the next VNF state $v_k^{t+1}$ is set as 1 (lines \ref{2line2}-\ref{2line3}). Otherwise, it proceeds to the next judgement. If the pending VNF state $v_k^{*t+1}$ is 1, the state keeps unchanged and the node resource utilization is updated (lines \ref{2line5}-\ref{2line6}). Then, the remaining SFCs that choose the same node are arranged based on the data size of SFCs in the ascending order and selected partly according to the remaining node resources (lines \ref{2line8}-\ref{2line9}).

\subsection{\textcolor{black}{Complexity Analysis}} 
Assuming that the width of the $i$-th layer of the neural network is $W_i$ and there exist $M$ layers in total, the computational complexity of forward propagation in Algorithm \ref{DDQN algorithm} is $\mathcal{O} (S\cdot A\cdot\sum_{i = 1}^{M-1} W_i W_{i+1} )$, where $S$ and $A$ are the numbers of elements in a state and an action, respectively. Besides, the complexity of Algorithm \ref{RepeatVNFState} is related to the number of nodes $\mathcal{I}$. There are $\mathcal{K}$ SFCs that need to be trained, and the total number of episodes and steps are $D$ and $P$, respectively. Hence, the total computational complexity is $\mathcal{O} (D\cdot P \cdot(K \cdot S\cdot A\cdot \sum_{i = 1}^{M-1} W_i W_{i+1} + \mathcal{I}))$.

\section{Simulation Results\label{sec:Simulation-Results}}

\begin{table}[!t]
    \renewcommand\arraystretch{1.3}
	\begin{center}
		\caption{PARAMETER SETTING} \label{parameter setting}
		\begin{tabular}{|c|c||c|c|}
			\hline
			Description & Value & Description & Value \\
			\hline
            \hline
			$P^{tr}_{g}$ & 0.5W &  $P^{tr}_u$ & 10W \\
            \hline 
            $\iota_0$ & 80dB & $h_u$ & 100m \\
            \hline
            $P_{uu}\&P_{us}$ & 10W & $f_{uu}$ & 2.4GHz\\
            \hline
            $\sigma_{uu}^2$ & $4\times 10^{-13}$W & $B_{uu}$ &4MHz\\
            \hline
            $P_{ss}$ & 20W & $G_{sg}^{tr}G_{sg}^{re}$ & 42dB\\
            \hline 
            $G_{us}^{tr}G_{us}^{re}$ & 42dB & $G_{ss}^{tr}G_{ss}^{re}$ & 52dB \\
            \hline
            $T_s$ & 1000K & $N_0$ & -114dBm\\
            \hline
            $B_{gu}$ & 2MHz & $B_{us}$ & 50MHz\\
            \hline
            $B_{ss}\&B_{sg}$ & 80MHz & $P_{sg}$ & 20W \\
            \hline
            $L_l$ & 2dB & $f^{cen}_{us}$ & 3.4GHz\\
            \hline
            $f^{cen}_{ss}$ & 2.2GHz & $f^{cen}_{sg}$ & 20GHz\\
            \hline
            $v_{n^t_i}$ & 12m/s & $P^{\rm MAX}_{n^t_i}$ & 5W\\
            \hline
            $N_0$ & -114dBm & $M_{n^t_i}$ & 0.5kg\\
            \hline
            $\mu_{n^t_i}$ & 20cm & $\nu _{n^t_i}$ & 4\\
            \hline
		\end{tabular}
	\end{center}
\end{table}

\subsection{\textcolor{black}{Simulation Setups}}
In this section, we conduct simulations for the SFC deployment in SAGIN using Python. The experimental hardware environment is based on the Intel(R) Core(TM) i9-10940X CPU, DDR4 64GB RAM, and GeForce RTX 3090 24GB*2 GPU. The specific parameters used in the simulation are listed in Table \ref{parameter setting}. The scenario is set up with 30 UAVs and 2 satellites. Among them, the UAVs are randomly arranged in a circle with a radius of 400 m and the distance between any two UAVs cannot be less than 20 m for safety consideration. The position information of the satellites is selected from the Starlink G6-35 near the coordinates of 32$^\circ$N, 119$^\circ$E on February 23, 2024 around 11:45 UTC. Compared with UAVs, satellites are more like auxiliary functional nodes, which can carry a larger amount of data. When UAVs are unable to carry the deployment and processing of tasks, tasks are uploaded to the satellites. Therefore, the number of satellites set is small. The number of SFCs is set as 200, with 2 or 3 VNFs to be processed in each SFC, and the data volume is [500 Mbit, 4,000 Mbit]. In RTEG, it is assumed that both UAVs and satellites are kept relatively stationary in a time slot, and the length of the time slot is set as 5 seconds.

\begin{figure}[!t]
	\centering
	\subfloat[]{\includegraphics[width=0.97\columnwidth]{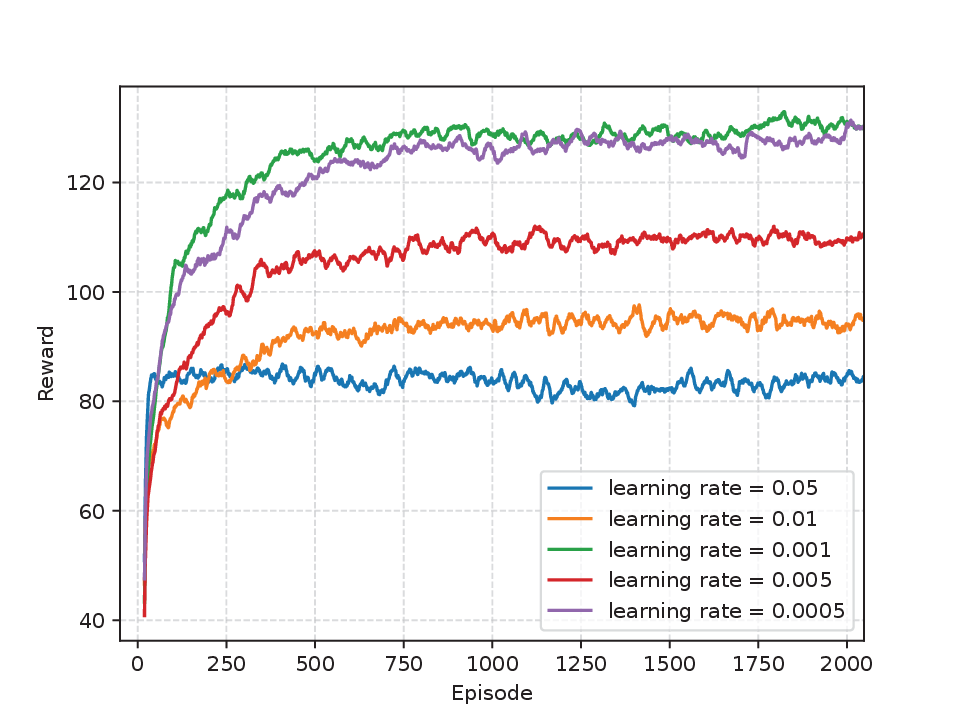}} \label{learningrate_reward}\\
	\subfloat[]{\includegraphics[width=0.97\columnwidth]{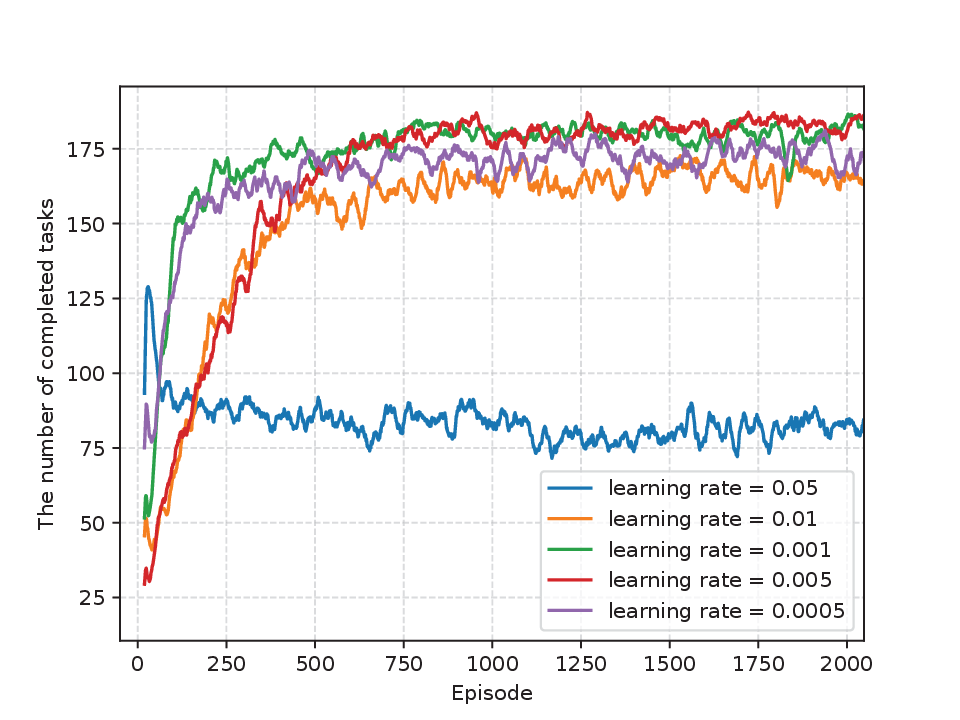}} \label{learningrate_task}
	\caption{The training effect under different learning rates. (a) Reward \textit{versus} episode. (b) The number of completed tasks \textit{versus} episode.}
    \label{fig:learningrate}
\end{figure}


During the simulation of DRL, the neural network structure consists of an input layer (the number of the state), three hidden layers (64, 32 and 32 neurons, respectively) and an output layer (the number of actions). The ReLU function is set as the activation function and the model parameters are updated using the Adam optimizer. In addition, we set the learning rate as 0.001, the discount factor as 0.9, and the $\epsilon$-greedy strategy in action selection is chosen linearly within [0, 0.9]. During optimizing the network model, an experience replay memory with capacity of 500 samples is set, and the selected batch size is 8. Then, a total of 3,000 episodes are performed, and the upper limit of step is set as 100, which is the maximum number of time slots in an episode.




\begin{figure}[!t]
	\centering
	\subfloat[]{\includegraphics[width=0.97\columnwidth]{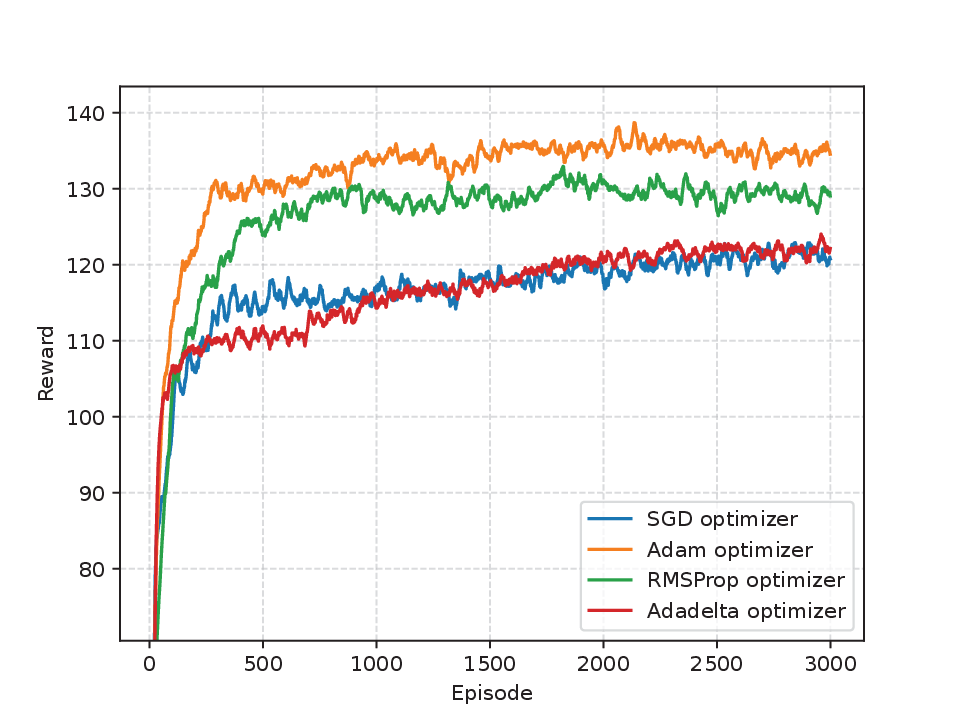}} \label{optimizer_reward}\\
	\subfloat[]{\includegraphics[width=0.97\columnwidth]{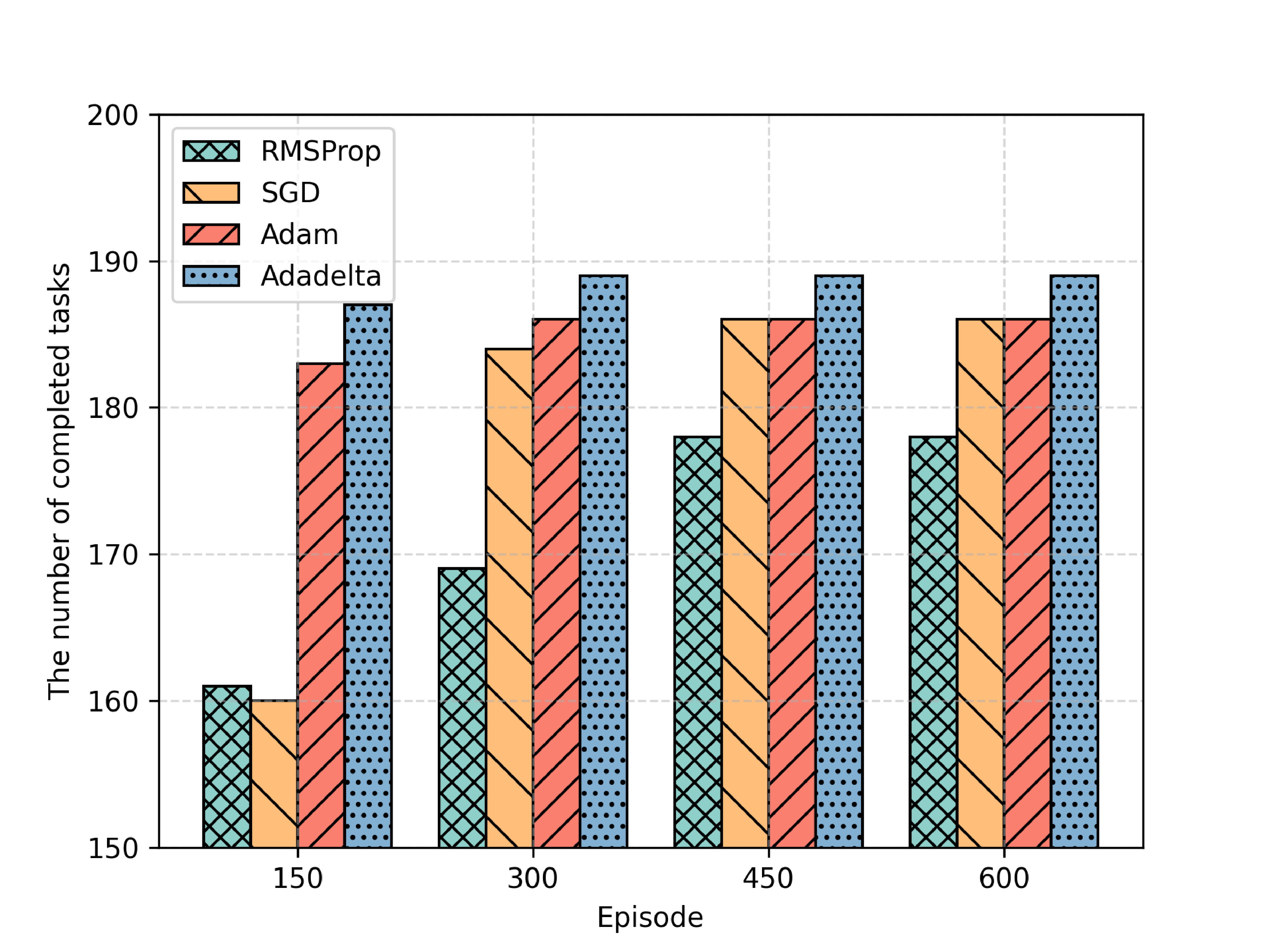}} \label{optimizer_task}
	\caption{Distinctions between different optimizers. (a) Reward \textit{versus} episode. (b) The number of completed tasks \textit{versus} episode.}
    \label{fig:optimizers}
\end{figure}

\subsection{\textcolor{black}{Simulation Results}}
\subsubsection{\textcolor{black}{Training Effect under Different Learning Parameters}}
The DRL-MSSNL-SAGIN algorithm is evaluated by different learning rates, as shown in Fig. \ref{fig:learningrate}. It is observed that various learning rates have different performances on the convergence of the algorithm as well as the results. At the beginning of the training, the rewards and results are unsatisfactory. With the increment of episodes, there is a significant rise in rewards, and the convergence speed is accelerated. The number of SFCs completing the deployment in limited time also gradually increases. When the learning rate is 0.001, the rewards and results are superior to other learning rates. If the learning rate is greater or less than 0.001, it shows different degrees of disadvantage. Among them, when the learning rate is 0.05, the results are the worst and do not converge to a more stable result, which means that a larger learning rate may lead to a local optimum rather than a global optimum. In addition, too small learning rate may cause DRL to be trapped in the local optimal solution, and can not jump out to find the global optimal solution. Due to the small step size, DRL may only explore around the local optimal solution in a small amplitude, and cannot conduct a broader search. Taking into account the actual implementation of the algorithm, the learning rate of 0.001 is selected.

Fig. \ref{fig:optimizers} shows the evaluation of the algorithm with different optimization strategies. It can be seen from Fig. \ref{fig:optimizers}a that the convergence speed of Adam optimizer is fast and it converges smoothly, while the rewards of Adadelta optimizer continue to grow and cannot converge quickly enough. Fig. \ref{fig:optimizers}b shows the completion of the optimization objective and it is obvious that the optimization results of these optimizers are different, in which the Adam optimizer relatively performs good. Thus, the Adam optimizer is chosen for the following simulations.

\begin{figure}[!t]
	\centering
	{\includegraphics[width=0.97\columnwidth]{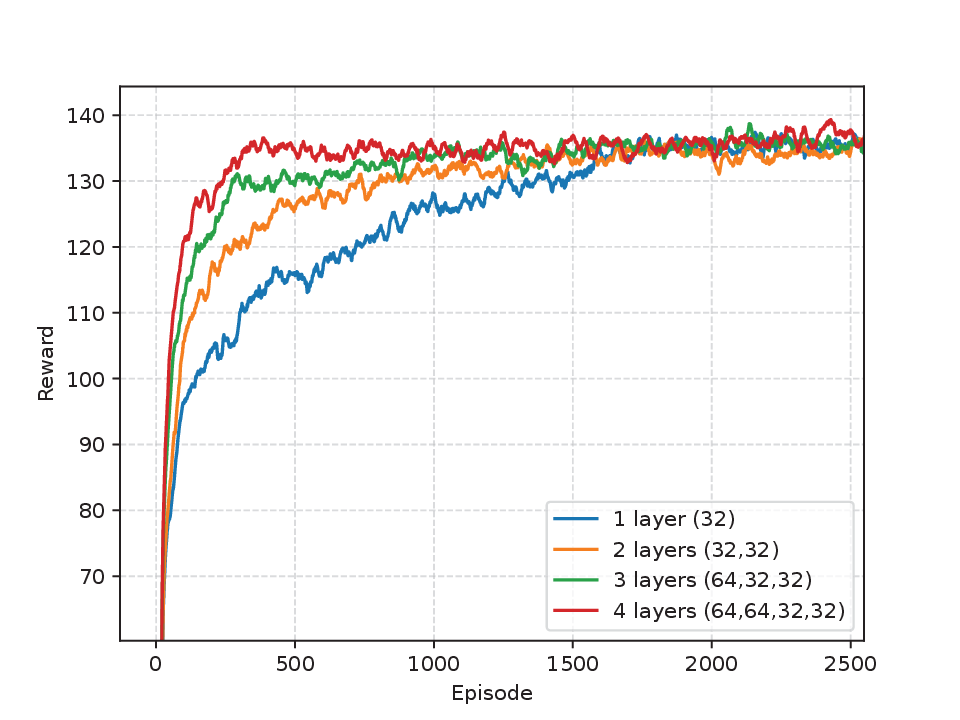}} \label{dnn_reward}
	\caption{The effect of convergence rate under different DNN layers.} 
    \label{fig:dnn}
\end{figure}

Different DNN structural layers also have impacts on the training situation of DRL, as shown in Fig. \ref{fig:dnn}. It is noted that when there is only one or two hidden layers, the convergence is slow. As the number of hidden layers increases, the training performs faster convergence. In addition, the results of three and four layers are similar, but four layers increase the training time. Therefore, to efficiently carry out the simulation, we leverage the three DNN structure layers (64, 32, 32).

\begin{figure}[!t]
	\centering
	\subfloat[]{\includegraphics[width=0.97\columnwidth]{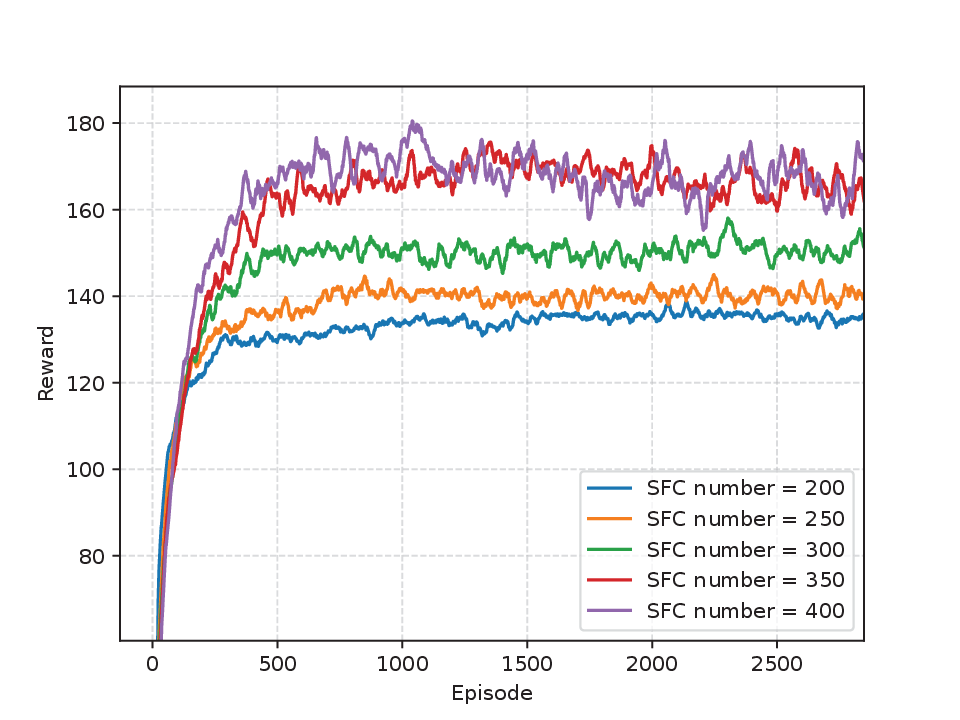}} \label{sfcnumber_reward} \\
	\subfloat[]{\includegraphics[width=0.97\columnwidth]{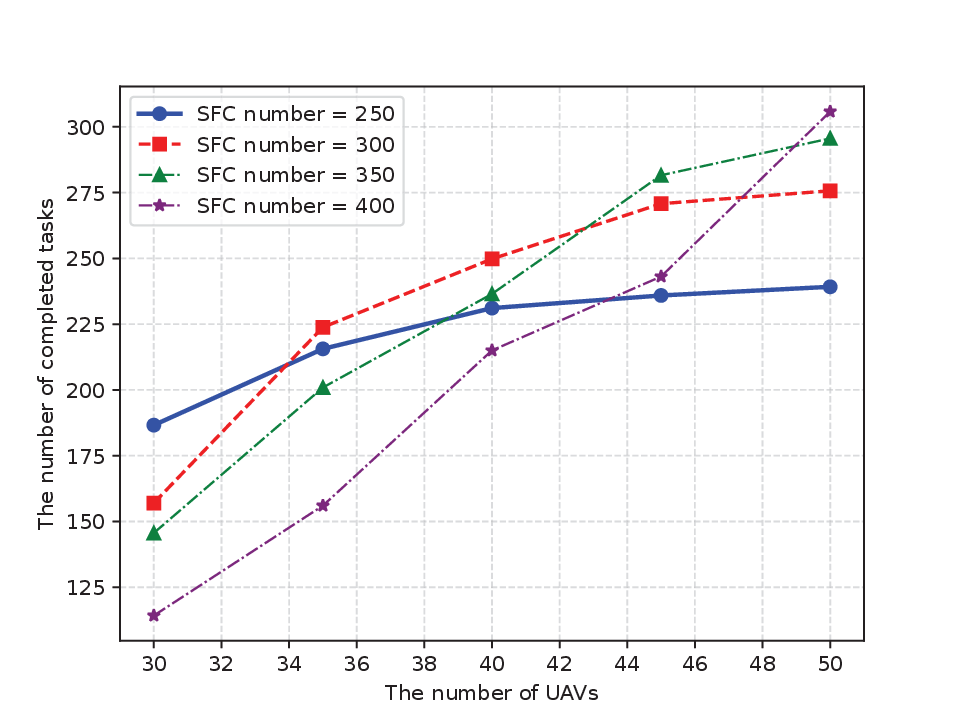}} \label{uavnumber_task}
	\caption{Performance comparison of different SFC numbers. (a) Reward \textit{versus} episode. (b) The number of completed tasks \textit{versus} the number of UAVs. }
    \label{fig:sfcnumber}
\end{figure}

Fig. \ref{fig:sfcnumber} shows the performance comparison of different SFC numbers, while other values are fixed. It is observed that with the increasing number of SFC, the completed number of SFCs is decreasing and fluctuating. Especially when the number of SFCs reaches 400 in Fig. \ref{fig:sfcnumber}a, it is difficult to have a stable convergence. It is accounted that the number of nodes and the corresponding resources are limited. The affordability of the whole network for various SFC numbers at different UAV sizes is compared in Fig. \ref{fig:sfcnumber}b. It is observed that when the number of SFCs is small, increasing the number of UAVs does not significantly impact the number of successful deployed SFCs. However, by combining Fig. \ref{fig:sfcnumber}a and Fig. \ref{fig:sfcnumber}b, if the SFC number reaches 400, the number of UAVs greatly impacts the outcome. When the UAV quantity grows, the number of successfully deployed SFCs also increases rapidly. It verifies that the SFC scheduling results are dependent on the scale of networks.

\begin{figure}[!t]
	\centering
	\subfloat[]{\includegraphics[width=0.97\columnwidth]{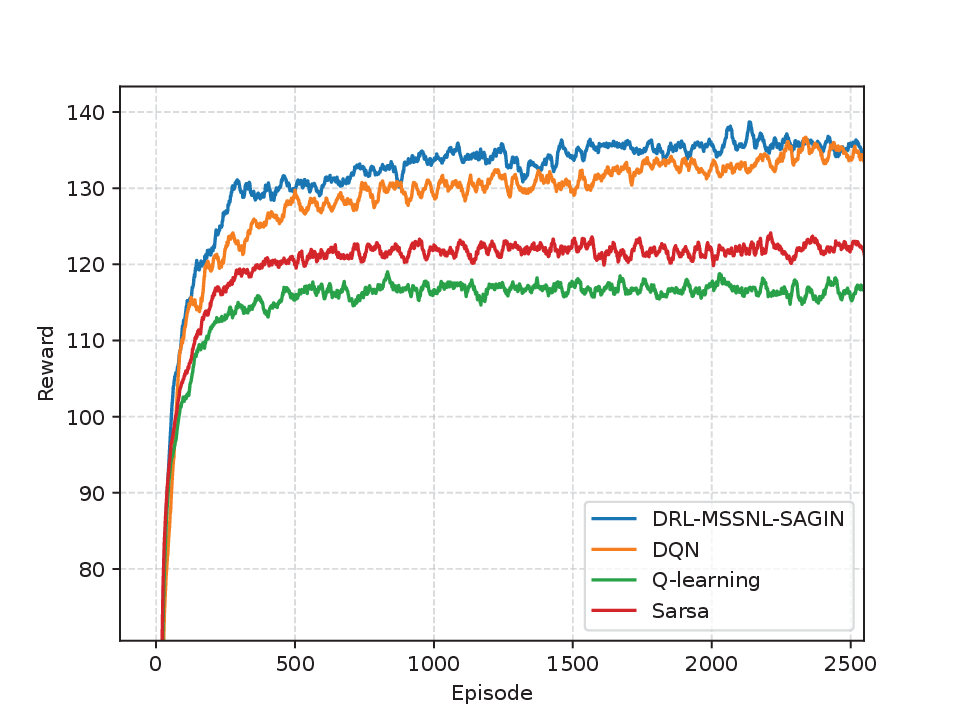}} \label{comparealg_reward}\\
	\subfloat[]{\includegraphics[width=0.97\columnwidth]{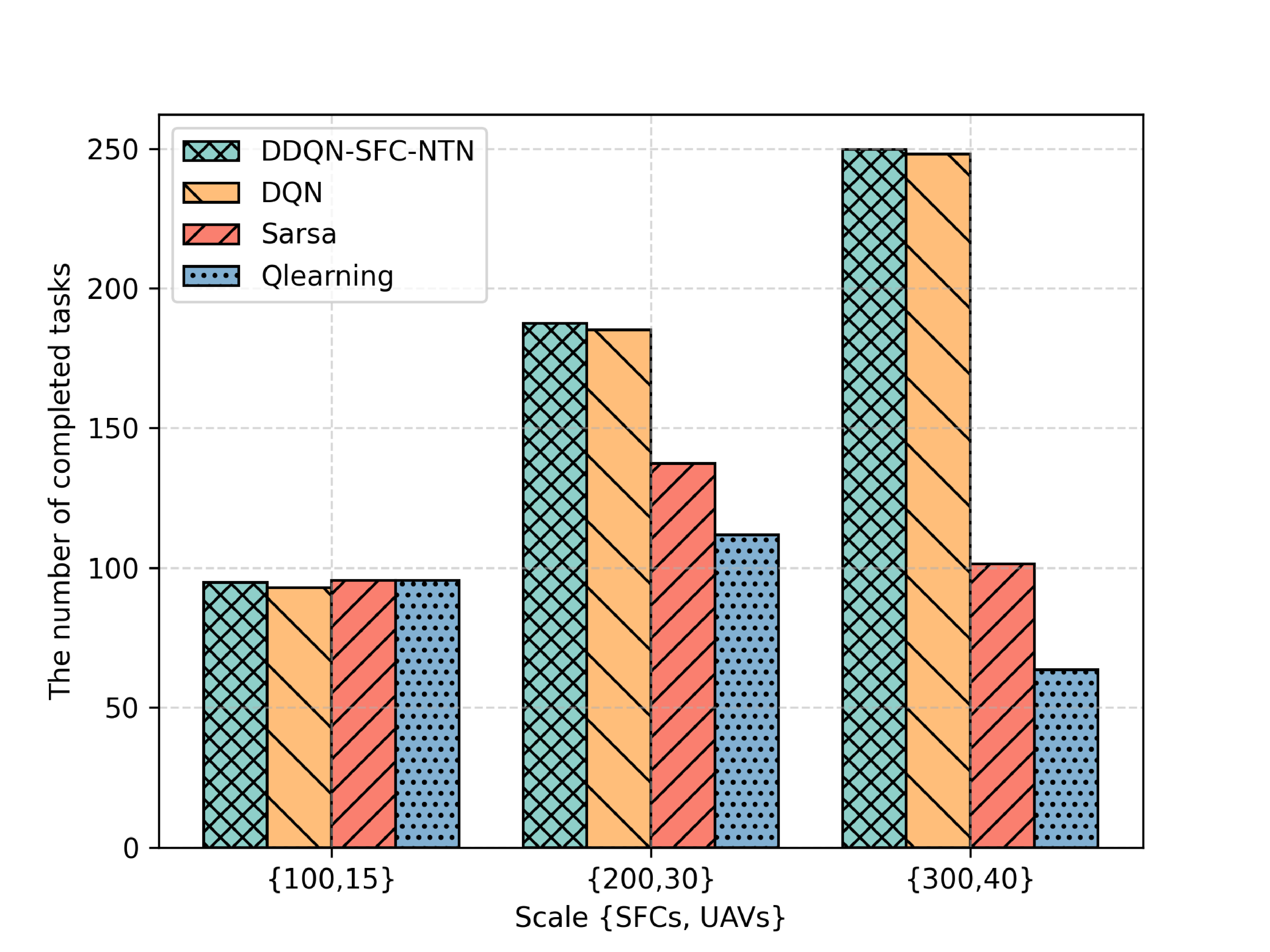}} \label{comparealg_scale}\\
    \subfloat[]{\includegraphics[width=0.97\columnwidth]{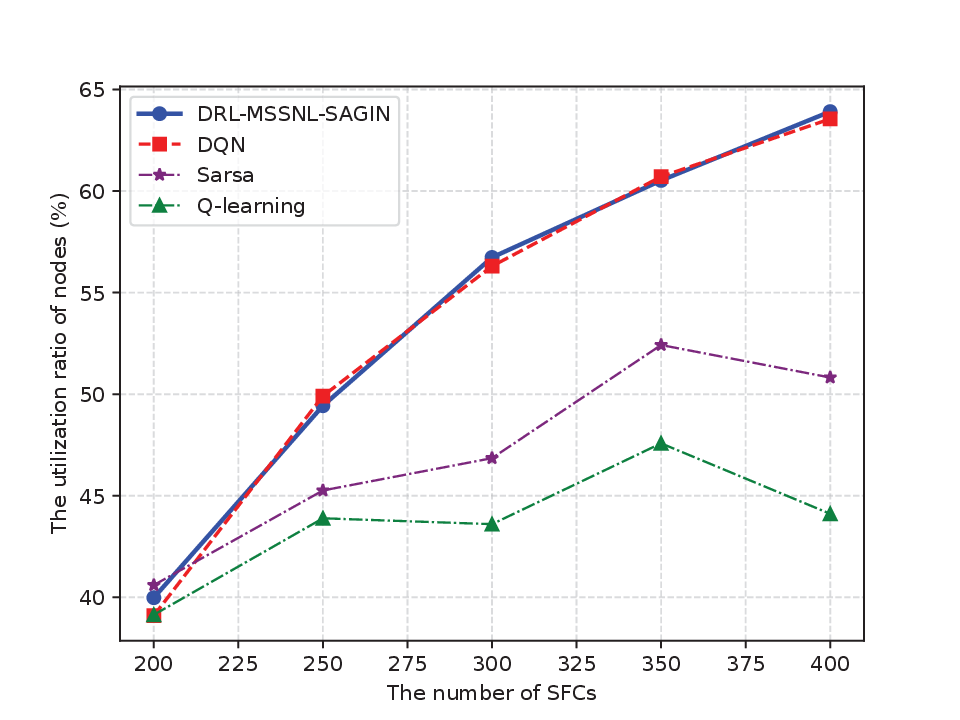}} \label{comparealg_task}
	\caption{Comparison of various RL algorithms. (a) Reward \textit{versus} episode. (b) The number of completed tasks \textit{versus} scale of SFCs and UAVs.} (c) The utilization ratio of nodes \textit{versus} the number of SFCs.
    \label{fig:comparealg}
\end{figure}

\subsubsection{\textcolor{black}{Comparison of Different Algorithms}}
We compare the proposed DRL-MSSNL-SAGIN algorithm for the SFC scheduling problem with Q-learning, Sarsa, and DQN, as shown in Fig. \ref{fig:comparealg}. It is observed in Fig. \ref{fig:comparealg}a that the convergence speed of four algorithms is similar, and when the number of SFC is small, the results are also similar. Moreover, in Fig. \ref{fig:comparealg}b, when the number of SFCs and UAVs increases, i.e., when the scales of networks and tasks increase, the results of Q-learning and Sarsa are unsatisfactory, since only a part of SFCs successfully deployed in limited time. Besides, there is a big gap between the results of Q-learning, Sarsa and DRL-MSSNL-SAGIN. Furthermore, Q-learning has the worst results, and Sarsa performs only slightly better than Q-learning. In addition, it is evident from Fig. \ref{fig:comparealg}c that the node resource utilization of Q-learning and Sarsa does not raise significantly with the increment of the SFC number. Hence, these two algorithms are completely unsuitable for large-scale network. On the contrary, the results of DRL-MSSNL-SAGIN and DQN are obviously better than Q-learning and Sarsa, and the proposed algorithm is slightly better than DQN, which is consistent with the characteristics of DDQN and DQN. When the number of SFCs increases, the resource utilization of nodes also increases significantly. Especially when the number of SFCs is increased to 400, the optimization result of DRL-MSSNL-SAGIN is more than twice of the results of Q-learning and Sarsa, and the resource utilization is approximately 15\% greater than Sarsa and 20\% greater than Q-learning. Therefore, the proposed  DRL-MSSNL-SAGIN algorithm has excellent performance for large-scale networks.


\section{Conclusions\label{sec:Conclusions}}

In this paper, we considered the highly dynamic characteristics of SAGIN and investigated the SFC deployment and scheduling model by proposing RTEG. The SFC scheduling problem was formulated with the objective of maximizing the number of successfully deployed SFCs in a finite time horizon, with the considerations of channel conditions, energy constraints, deployment limitations, and multiple resource capacities. To tackle this problem, we reformulated it as an MDP and proposed the DRL-based algorithms. Besides, we designed the algorithm for VNF state transition to achieve the efficient SFC scheduling. Via simulations, we analyzed the influences of various parameters on the SFC scheduling problem, and selected the appropriate parameters to obtain better optimization results. Simulations also showed that the proposed algorithm outperformed other benchmark algorithms, with respect to fast convergence, better optimization results, and efficient resource utilization.


\textcolor{black}
{\bibliographystyle{IEEEtran}
\bibliography{ref}
}

\end{document}